\documentclass[journal]{IEEEtran}

\usepackage[pdftex]{graphicx}
\usepackage{lscape}
\usepackage{xcolor}
\usepackage{tikz}
\usepackage{amsmath}
\usepackage{amssymb}
\usepackage[version=4]{mhchem}
\usepackage{chemfig}
\usepackage{booktabs}
\usepackage{comment}
\usepackage{textcomp}

\def\checkmark{\tikz\fill[scale=0.4](0,.35) -- (.25,0) -- (1,.7) -- (.25,.15) -- cycle;} 

\begin{document}

\title{Molecular Communications in Viral Infections Research: Modelling, Experimental Data and Future Directions}

\author{Michael Taynnan Barros$^*$, Mladen Veleti\'{c}$^*$, Masamitsu Kanada, Massimiliano Pierobon, Seppo Vainio, Ilangko Balasingham, Sasitharan Balasubramaniam
\thanks{Authors with $^*$ contributed equally to the manuscript.}
\thanks{M. T Barros is with the School of Computer Science and Electronic Engineering, University of Essex, Colchester, UK, and also with the CBIG/BioMediTech, Tampere University, Tampere, Finland.}
\thanks{M. Veleti\'{c} is with the Intervention Centre, Oslo University Hospital, Oslo, Norway, and also with Department of Electronic Systems, Norwegian University of Science and Technology, Trondheim, Norway.}
\thanks{M. Kanada is with the Institute for Quantitative Health Science and Engineering, Michigan State University, East Lansing, MI, USA.}
\thanks{M. Pierobon is with the Department of Computer Science and Engineering, University of Nebraska-Lincoln, Lincoln, NE, USA.}
\thanks{S. Vainio is with the Biocenter Oulu, Laboratory of Developmental Biology, Oulu University, Oulu, Finland.}
\thanks{I. Balasingham is with the Intervention Centre, Oslo University Hospital, Oslo, Norway, and also with Department of Electronic Systems, Norwegian University of Science and Technology, Trondheim, Norway.}
\thanks{S. Balasubramaniam is with the Telecommunications Software and Systems Group, Waterford Institute of Technology, Waterford City, Ireland.}
}

\markboth{Submitted for journal publication}%
{Shell \MakeLowercase{\textit{et al.}}: Bare Demo of IEEEtran.cls for IEEE Journals}

\maketitle

\begin{abstract}
Hundreds of millions of people worldwide are affected by viral infections each year, and yet, several of them neither have vaccines nor effective treatment during and post-infection. This challenge has been highlighted by the COVID-19 pandemic, showing how viruses can quickly spread and how they can impact society as a whole. 
Novel techniques that bring in different disciplines must emerge to provide forward-looking strategies to combat viral infections, as well as possible future pandemics. In the past decade, an interdisciplinary area involving bioengineering, nanotechnology and information and communication technology (ICT) has been developing, known as Molecular Communications. This new emerging area uses elements of classical communication systems, and maps it to molecular signalling and communication found inside and outside the body, where the aim is to develop new tools that can serve future medicine. One of these tools is the ability to characterise the signalling processes between cells and infectious disease locations at various levels of the human body. In this paper, we provide an extensive and detailed discussion on how Molecular Communications can be integrated into the research on viral infectious diseases modelling, and how possible treatment and vaccines can be developed considering molecules as information carriers. We provide a literature review on the existing models of Molecular Communications for viral infection (in-body and out-body), a deep analysis on their effects on the host and subsequent communication process for other systems within the body (e.g., immune response), sources of experimental data on known viral infections and how it can be used by the Molecular Communications community, as well as open issues and future directions. Since the development of therapeutics/vaccines needs an interdisciplinary approach centred around ICT, we are confident that Molecular Communications can play a central role here by providing a detail characterisation and manipulation of the propagation of molecules in different media. 
\end{abstract}


\IEEEpeerreviewmaketitle

\section{Introduction}



The COVID-19 pandemic shocked the world by demonstrating the severity of the viral infection and how it can disrupt society by impacting human health as well as global economies. As of September 2020, more than 34 million people has contracted the disease resulting in just over a million deaths. Therefore, this particular virus has a mortality rate of approximately 4\%. During the first months of the pandemic, global stock markets experienced their worst crash since 1987, in the first three months of 2020 the G20 economies fell by 3.4\% year-on-year, an estimation of 400 million full-time jobs were lost across the world, and income earned by workers globally fell 10\%, where all of this effects is equivalent to a loss of over US\$3.5 trillion \cite{jones2020coronavirus}. As a result, governments around the world have quickly formulated new recovery plans, where for example in the EU, an investment of 750 billion euros is set to bring the continent back to normality within the first half of the decade ((this also includes funding for research on COVID-19)) \cite{Matina2020coronavirus}. Despite these investments, the world must prepare for not only coping with this new disease and its various effects on the human health, but seeking for novel technologies that can help minimise, or even block, future pandemics.


The SARS-CoV-2 virus itself is likely to remain a challenge for the next couple of years despite the development of vaccines \cite{saxena2017mesenchymal}. First, it is challenging to develop a vaccine that is effective for different virus strains and their mutations. 
Besides, for patients that are infected, the detrimental effect of the virus in the human body can leave lifelong consequences to tissues and organs. To give an example of the difficulty of eradicating viruses, historic virus such as influenza had its first pandemic in the 16$^{\text{th}}$ century and is still considered a global health challenge till this present day \cite{thompson1852annals}.
Therefore, constant efforts in new robust vaccines, as well as drugs, are continuously being sought and this requires the development of new technologies that focus on the mechanisms of infections, and in particular the virus molecular relationships with the host cells \cite{roldao2010virus}. 


In the past 10 years, an interdisciplinary research area known as Molecular Communications has been developing, and it bridges the areas of communication engineering and networking, molecular biology, as well as bioengineering \cite{Akyildiz2019,nakano2017molecular}. This area focuses on realising radical new technology for a society that holds the promise for subtle sensing and actuation capabilities inside the human body through a network of micro- and nano-sized devices \cite{akyildiz2015internet,akyildiz2008nanonetworks}. These devices can use the existing natural signalling of cells and tissues to interact, as well as communication with the human body. The main advantage is the ability to increase the biocompatibility of implantable systems, and to achieve this through the integration of synthetic biology \cite{akyildiz2015internet,akyildiz2008nanonetworks}. 
This novel research area can have a central role to combat current and future pandemics, not only for understanding new insights into the viral properties and characteristics, but also for novel treatments \cite{felicetti2016applications,atakan2012body,barros2018multi,veletic2019synaptic}. Molecular Communications can contribute (a) to the characterisation of the virus propagation within the body, (b) to understand the mechanism used by the virus to enter the human body, or mechanism of expulsion, and (c) to understand how the airborne virus propagates in the air.
Additionally, although not covered in this survey, is the paradigm of Internet of Bio-Nano Things \cite{akyildiz2015internet}, where Molecular Communications can facilitate the communication between engineered cells for viral infection detection/therapy, and Bio-Cyber Interfaces that can transmit the data to cloud-based digital healthcare services \cite{balasubramaniam2012realizing}. 

However, the literature in Molecular Communications does not provide a wide range of work that tackles the issue of viral infection as a whole. Even though there are models for Molecular Communications for bacterial infection \cite{martins2018molecular,martins2016using}, there have not been any surveys proposed for Molecular Communications models of viral infections. 
To date, Molecular Communications models for viral infection includes multi-hop transfer of genetic content through diffusion over extracellular channels \cite{walsh2013reliability}, viral propagation in the air \cite{khalid2019communication,khalid2020modeling}, propagation within the respiratory tract \cite{vimalajeewa2020silico}, as well as interactions with host cells \cite{martins2018computational}. Even though these models are very interesting and provide a formidable representation of biology through the glasses of a Molecular Communications researcher, the issue of virus propagation and the infection itself are much more complex. First, models must gather all necessary information about the infection process, which comprises of the replication of viruses and in-body propagation, going down to the interaction of genes and host cells, as well as the virus spike proteins effects to binding host cell receptors. Secondly, virology is a very active research area and has collected many resources over the years (data and tools) that can benefit Molecular Communications research. In order to develop research work with a strong societal impact in order to tackle viral infectious diseases, Molecular Communications researchers are required to bridge the gap between communication theory and experimental biology, and in particular the use of available data. 

This paper presents a literature review and analysis of existing models and data for Molecular Communications. The goals of this paper are as follows: 1) to provide how the infectious disease is currently modelled using Molecular Communications; 2) to provide a deep analysis on the existing models to provide a direction on how they should be improved, looking from a biological standpoint; 3) to provide initial guidelines on what experimental data can be used and how they should be integrated to Molecular Communications models, and 4) to identify the main challenges and issues that the community should focus on moving forward. We recognise that Molecular Communications can support not only the understanding of infectious diseases, but it can also elicit the development of novel technologies for both sensing and actuation in the body based on how viruses propagate, are transmitted and received by host cells. 

The paper has the following contributions:
\begin{itemize}
    \item \textit{A literature survey on models of infectious diseases for in-body and out-body Molecular Communications:} We present a deep analysis on existing Molecular Communications models looking at how to address the issue of bridging these models closer to the existing biological literature and data on infectious disease. For the in-body models, we investigate the virus entry mechanism, the virus spread and the immune system response. For the out-body models, we look into the transmitter, channel and receiver processes for human-human propagation of infection.
    \item \textit{Analysis of existing open data on viral infections that can be utilised by Molecular Communications researchers:} We collect a variety of data from a number of sources that we believe can be used by the community to gain a better understanding of viral infections. These data vary from the genetic information of a variety of viruses to the molecular structure and the effects on the hosts, e.g., the immune markers and host infection impact. We focus on a selection of viruses that includes the SARS-CoV (1-2), MERS-CoV, Ebola (EBOV), Dengue (DENV), Zika (ZIKV) and hepatitis C (HCV).
    \item \textit{Overview on open issues and challenges:} Based on the many opportunities for research in Molecular Communications and infectious diseases, we provide five different points where we present a deep analysis aiming to highlight what are the main topics to drive future research on Molecular Communications. We include discussion about: 1) linking experimental data to Molecular Communications models, 2) novel in-body viral intervention techniques, 3) emerging technologies for infection theranostics (therapy and diagnostics), 4) bridging Molecular Communications and bioinformatics tools, and 5) novel Molecular Communications models.
\end{itemize}

The paper structure is as follows. Section \ref{sec:background} presents the background information on infectious disease. Section \ref{sec:molecom} presents a literature survey on viral models for infections in in-body and out-body settings. Section \ref{sec:expdata} presents a set of experimental data that can be exploited in Molecular Communications. Section \ref{sec:open} presents the open issues and challenges for the future of Molecular Communications research on communicable disease. Finally, Section \ref{sec:conclusions} concludes the study.

\section{Background Information on Infectious Diseases} \label{sec:background}

In this section, we go through several known communicable viral disease and provide examples of devastating outbreaks in the 21$^{\text{st}}$ century. We select seven viruses that do not have a licensed vaccine for treatment or where the intervention mechanisms are only used to alleviate symptoms of the hosts. Our survey focuses on three families of viruses, and this includes \emph{Coronavirus}, \emph {Filovirus}, and \emph{Flavivirus}. 


\subsection{Coronavirus}
\textit{Coronaviridae} is a family of viruses that includes \textbf{SARS-CoV-1}, \textbf{MERS-CoV} and \textbf{SARS-CoV-2}, and contains single-stranded positive-sense RNA of the size of 31 kilobases (kb). The most severe virus in this family is the SARS-CoV-2. Example properties of SARS-CoV-2 include asymptomatic infection to severe pneumonia and replicates through a variety of cells that exhibit \emph{Angiotensin Converting Enzyme 2} (\emph{ACE2}) expression (a number of these cells are found in the respiratory tract, and in particular deep in the alveolar regions). SARS-CoV-1 and MERS-CoV are known to cause severe pneumonia with high replications rates in the respiratory tracts. The immune response of the three different viruses is also very different. In the case of SARS-CoV-1 and MERS-CoV, antibodies response are at an early stage of the infection process. However, this is not the case for SARS-CoV-2, where the symptoms from the infection process can take up to two weeks.

There are several differences between the SARS-CoV-1, MERS-CoV and SARS-CoV-2 in the spreading process. While SARS-CoV-1 and MERS-CoV are known to develop severe pneumonia, they have exhibited limited person-to-person spreading, which is very different from SARS-CoV-2 \cite{Sariol20}. Even though promising solutions for vaccines targeting SARS-CoV-2 are in the testing phase, their efficacy is yet unknown or unpredictable. Besides vaccines, the existing immune-based treatment (e.g., plasma transfusion) is only found to have temporary effects \cite{florindo2020immune}. On top of that, there are other several unknowns about how SARS-CoV-2 affects different organs, indicated by clinical data \cite{lebeau2020deciphering,wadman2020rampage}. 

\subsection{Filovirus}
\textit{Filoviridae} family of viruses comes from the thread-like structure of the virus that also contains many curvy branches \cite{Ansari14}. The most common, or well known, is the \textbf{EBOV} (\textit{zaire ebolavirus}). The size of the EBOV is approximately 970 nm--1200 nm, and has a diameter of 80 nm. The virus contains seven genes that encode for nucleoprotein, glycoprotein, RNA dependent RNA polymerase, and VP24, VP30, VP35 and VP40, which are the structural proteins. In the structure, the virus contains an outer envelope, and this covers the middle layers of VP40 and VP24 that protect the virus genome. This enables the virus to enter the cells in order to conduct replication. Currently, there is only one FDA approved Ebola vaccine (approved in 2019 \cite{us2019first}) that has a successful performance around 70\% to 100\% efficiency, and this is the rVSV-ZEBOV vaccine \cite{henao2017efficacy}. This vaccine acts as the glycoprotein duplicate, where once expressed in the host, it will activate the immune system response. The vaccine was used for clinical trials in West-Africa, to cope with the local 2016 pandemic.
However, research is still on-going to analyse the vaccine response for virus genome mutation.

\subsection{Flavivirus}
\textit{Flaviviridae} is a family of viruses that is mainly characterised by the yellow complexion found on the hosts after infection (hence yellow fever), and by the transmission mode of arthropod vectors (mainly ticks and mosquitoes). We analyse three types of virus in this family, and they include \textbf{DENV}, \textbf{ZIKV} and \textbf{HCV}.

The single positive-stranded RNA DENV is mosquito-borne and mostly found in countries in the centre global hemisphere, where the warm temperature is an ideal location for mosquito's habitation. The virus has not always been found to transmit through mosquitoes. Many years ago, the transmission mode was sylvatic, meaning contraction from wild animal contacts. Over the past 20 years, dengue fever has increased dramatically, affecting more than 390 million people each year. The main molecular characteristic of this virus is the genomic RNA surrounded by numerous protein layers. There are three, so-called, structural proteins: capsid protein $C$, membrane protein $M$, and envelope protein $E$. This is one of the leading causes in the difficulty of discovering an appropriate vaccine, alongside the five different serotypes, which remains a medical challenge \cite{normile2013surprising}. The virus also contains seven non-structural proteins (NS1, NS2a, NS2b, NS3, NS4a, NS4b, NS5). The non-structural proteins create a strong response within the host immune system, with some studies pointing to an actual increase in transmission and infection worsening when it tries to fight the virus off \cite{conway2016aedes}. 

\begin{figure*}[t!]
    \centering
    \includegraphics[width = \linewidth, trim=0cm 8.75cm 0cm 0cm]{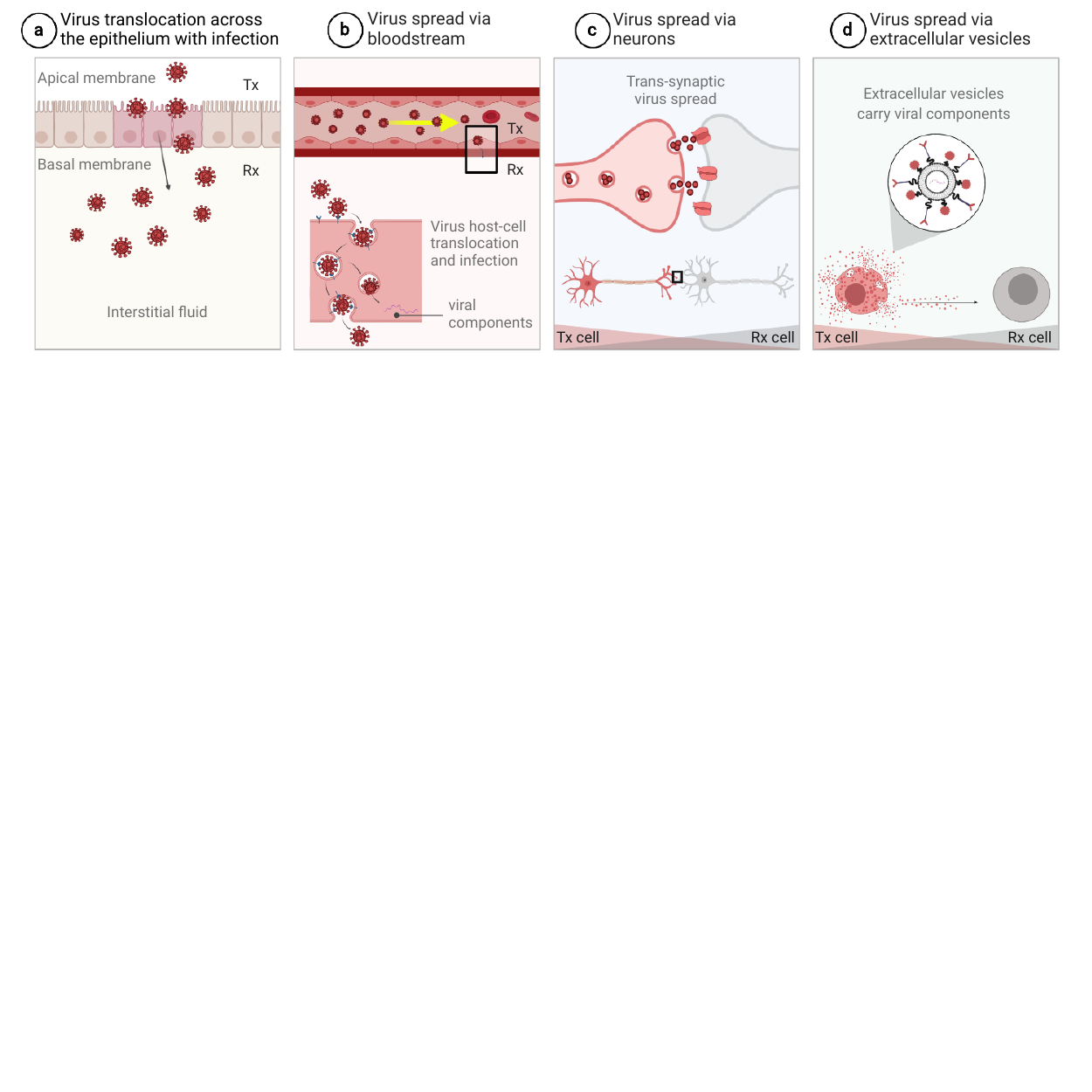}
    \caption{Molecular Communications channels of viral in-body spread. After (a) translocation across the epithelium, the virus spread throughout the body utilising (b) the circulatory system, (c) nervous network, and (d) cell-released EVs that carry viral components as their cargo and deliver to other cells, eventually causing systemic infection.}
    \label{fig: ViralSpread}
\end{figure*}

ZIKV is transmitted and similarly affects the host cells to the DENV since both share a distinct genetic component. However, the fever from ZIKV infection is more potent and is found to impact developing fetus in pregnant women, and can lead to microcephaly. This virus is relatively new. Hence no vaccine is available. Similar to the DENV, there are seven non-structural proteins, three structural proteins and a positive single-stranded RNA genome. However, the main difference is the mechanism that the host cells reacts to the genome upon infection, where the infected cells are found to progress into the swollen stages, and this leads to cell death. This is only possible through the virus gene expression and the inability of the host cell to protect itself against virus binding through the concentration of the IFITM3 protein.

Lastly, HCV is also a single-stranded RNA virus that primarily affects the liver and its functions. Infected hosts can have symptoms that include occasional fever, dark urine, abdominal pain and yellow-tinged skin. To date, the virus has infected nearly 71 million people worldwide. The virus itself is simpler than the other virus in the family, but still have more than one structural proteins, which are the E1 and E2 glycoproteins. They are not as effective as more complex structural proteins,
but they can be considered efficient in shielding the virus against the host immune system. They also have six non-structural proteins (NS2, NS3, NS4a, NS4b, NS5a, NS5b). Even though simpler in structure due to its numerous genotypes derived from their protein structure, there is still no vaccine for HCV, or even proper medical intervention techniques. These efforts would help minimise damaging effects to the liver, the organ most damaged by this virus in infected patients \cite{yu2010new}.

\section{Molecular Communications for Viral Infections} \label{sec:molecom}

In this section, we present the literature review on the in-body- and out-body Molecular Communications models.



\subsection{In-body Molecular Communications Models} \label{sec:inbody}
\subsubsection{Virus Entry Mechanisms}
The Molecular Communications paradigm gives us a clear understanding of how the virus acts and distributes within the body over time. In the context of communications, the virions are considered as information carriers, which propagate messages (genome) from the location of transmission until the location of the reception, which can be the host cells in specific organs or tissues. The information conveyed by the virions is the \textbf{infection action}.

In theory, a single virion is enough to enter the body and initiate a viral infection, provided that the host cells are accessible to the viral binding process. Besides the accessible cells being \textit{susceptible} to infection -- they must also express the receptors to which the virus binds, and \textit{permissive} to infection -- which means they must contain protein and machinery necessary for virus replication~\cite{Louten2016}.  

If the virus enters the host through the \textit{respiratory tract}, \textit{gastrointestinal tract}, \textit{genital tract} or \textit{optical tract}, the main barrier between the virus and internal environment of the body is the \underline{epithelial cells} -- the layer of cells that line the outer surfaces of organs and blood vessels and the inner surfaces of cavities (Fig.~\ref{fig: ViralSpread}a). The epithelial cells of the respiratory tract are targeted by SARS-CoV (1-2) and MERS-CoV viruses as the most common portal of entry. Unlike SARS-CoV (1-2), which exclusively infects and releases through the apical route\footnote{The apical membrane faces the external (luminal) compartment and contains proteins that determine secretion and absorption, whereas the basolateral domain faces the internal (systemic) compartment (tissues and blood).}, MERS-CoV can spread through either side of human bronchial epithelial cells. SARS-CoV (1-2) and MERS-CoV viruses contained in larger droplets are deposited in the upper respiratory tract (the nose, nasal passages, sinuses, pharynx, and larynx), while smaller aerosolised particles or liquids are transferred into the lower respiratory tract (the trachea, bronchi, and lungs). EBOV targets the epithelial cells as a final attack though, after infecting fibroblasts of any type (especially fibroblastic reticular cells), mononuclear phagocytes (with dendritic cells more affected than monocytes or macrophages) and endothelial cells. On the other hand, if the virus is delivered through penetration of the \textit{skin} (e.g., DENV- or ZIKV infection from a mosquito bite), \textit{wounds} or \textit{transplantation} of an infected organ (e.g., HCV infected organ), the epithelium is bypassed. 

Viruses have evolved strategies to translocate across the epithelial barrier and act as pathogens. They can enter and infect or cross epithelial cells through the following three modes~\cite{Bomsel2003}: 1) Endocytosis and transcytosis (without infection), 2) Polarised surface entry and infection by fusion, and 3) Endocytosis and endosomal fusion with infection.

\textit{Endocytosis and transcytosis (without infection)} are, respectively, entry- and intracellular transport mechanisms for specific viruses, such as poliovirus, reovirus and human immunodeficiency virus 1 (HIV-1), performed by specific lymphoid areas of the gastrointestinal tract covered by specialised epithelial cells known as M cells. During endocytosis, which is initiated at clathrin- and caveolin-coated pits and vesicles, or lipid raft microdomains, the host cell engulfs the virus. During transcytosis, the host cell transports the virus through its cytosol and eventually eject the virus at the opposite side of the membrane. 
\textit{Polarised surface entry and infection by fusion} is an entry mechanism for enveloped viruses, including SARS-CoV (1-2) and MERS-CoV, whose genome is surrounded by a capsid and a membrane~\cite{Cong2014}. The virus fuses either to the apical membrane or the basal membrane of the epithelial cell and transfers the genome into the cytoplasm. 
Lastly, \textit{endocytosis and endosomal fusion with infection} is an entry mechanism for both enveloped and naked (the genome is surrounded only by a capsid) viruses, including EBOV. Other examples include influenza virus types A and C, bovine coronavirus, hepatitis A (HAV), vesicular stomatitis virus (VSV), primary herpes simplex virus (HSV), human cytomegalovirus (HCMV), adeno-associated virus (AAV)-2, simian virus 40 (SV40), measles virus, Semliki Forest virus (SFV), Sindbis virus, Jamestown Canyon (JC) polyomavirus, parvovirus, and the minor group of human rhinoviruses (HRV). These viruses internalise and retain in transport vesicles. To gain access to the cytoplasm, their genome has to leave the vesicle by which it was taken up, usually by penetrating the host cell cytosol through fusion from an endosome. 

We refer to modelling viral translocation across the epithelial barrier as \textsc{\textbf{Model 1}}. Despite all differences in the mechanisms involved in this model, 
the transfer process always starts with the virions binding to the target receptors. Upon binding, the virions become fused with- or internalised into the host cell cytosol. Recycling/negative feedback mechanisms regulate the number of surface bonds between the virions and the receptors. This leads to the following chemical kinetic model representing the viral load/concentration at the extracellular space $V_o(t)$, the viral load at the epithelial host cell membrane (host cell-bound virions) $V_b(t)$, and the viral load in the host cell cytosol (fused or internalised virions) $V_i(t)$, respectively~\cite{Veletic2019}:
\begin{align}
\beta \frac{\mathrm{d}V_o(t)}{\mathrm{d}t} =& \beta \left[V_{in} - cV_o(t)\right] - \nonumber \\ -& aV_o(t)\left[n_vN(t) + n_vN_0(t) - V_b(t)\right] \label{e: mass-action6}\\
\frac{\mathrm{d}V_b(t)}{\mathrm{d}t} =& aV_o(t)\left[n_vN(t) + n_vN_0(t)-V_b(t)\right]-k_iV_b(t) \label{e: mass-action7}\\
\frac{\mathrm{d}V_i(t)}{\mathrm{d}t} =& k_iV_b(t) \label{e: mass-action8},
\end{align}
where $\beta$ is the ratio of the volume of the considered medium containing $V_o(t)$ and the host cell volume, $V_{in}$ is the initial viral load at the extracellular space, $c$ is a constant rate of viral clearance per virion by mechanisms such as immune elimination (corresponding to a virion half-life $t_{V_{1/2}}=\ln(2)/c$), $n_v$ is the total number of the viruses that can be bound, $N(t)$ is the total number of occupied receptors per unit volume at the membrane, $N_0(t)$ is the total number of unoccupied receptors per unit volume at the membrane, $a = a_0/n_v$ is the rate defined through the maximal binding rate $a_0$ measured when none of the viruses is bound to the membrane, and $k_i$ is the virus fusion or internalisation rate~\cite{Veletic2019}. 

The presented model can be considered accurate if the data of the concentration of viral ligands interaction with the host cell receptors are available.

\subsubsection{Virus Spread}
After translocation across the epithelial barrier, the virus infects and replicates at the site of infection, causing \textbf{localised infections}, and/or initiates infection through one organ and then spreads to other sites, causing \textbf{systemic infections}~\cite{Louten2016}. 

A straightforward way to describe the viral load ($V^{(l)}(t)$) dynamics in a localised infection is to use the target cell-limited model~\cite{Zitzmann2018}. This model neglects intracellular processes and takes into account uninfected susceptible target cells ($T$) and infected virus-producing cells ($I$) within an observed organ. The basic model is formulated by the following system of nonlinear ordinary differential equations (ODEs)~\cite{Bonhoeffer1997}: 
\begin{eqnarray}\label{e: virusLocal}
\frac{\mathrm{d}V^{(l)}(t)}{\mathrm{d}t}= pI(t) - cV^{(l)}(t),
\end{eqnarray}
where
\begin{eqnarray}
\frac{\mathrm{d}I(t)}{\mathrm{d}t} &=& kV^{(l)}(t)T(t) - \delta I(t),\\
\frac{\mathrm{d}T(t)}{\mathrm{d}t} &=& \lambda - dT(t) - kV^{(l)}(t)T(t) \label{e: targetLocal}.
\end{eqnarray}
The target cells become infected cells which produce virus with production rate $p$, $k$ is a constant infectivity rate, $\delta$ is a constant rate of death in infected cells (corresponding to an infected cell half-life of $t_{I_{1/2}}=\ln(2)/\delta$), $\lambda$ is a constant rate of uninfected target cells production, and $d$ is a constant rate of uninfected target cells death (corresponding to a target cell half-life of $t_{T_{1/2}}=\ln(2)/d$). This model can be applied to analytically describe the local spread of any family of viruses. Apart from the viruses considered in Section~\ref{sec:background}, rhinovirus and papillomavirus are examples of viruses that cause only a localised infection. Rhinovirus infects the epithelial cells of the upper respiratory tract and replicates there, whereas papillomavirus infects the skin and replicates in the epidermis. 

Describing the viral load dynamics in a systemic infection is more challenging since the virus spreads to other organs using mechanisms like the bloodstream (\textit{hematogeneous spread}), neurons (\textit{neurotropic spread}), or extracellular vesicles.

Viruses can enter the \textbf{bloodstream} either directly through inoculation into an animal or insect bites (e.g., DENV and ZIKV), or through the release of virions produced at the entry site into the interstitial fluid (e.g., coronaviruses)~\cite{Louten2016}. This fluid can be taken up by lymphatic vessels that lead back to lymph nodes. Although immune system cells filter the interstitial fluid within the lymph nodes, some virions escape immune cells and continue within the interstitial fluid, which is eventually returned to the bloodstream. 
The virus takes advantage of the blood distribution network for the propagation of the virions from a location they are injected into the blood flow to a targeted site within reach of the cardiovascular system. Advection and diffusion are the mass transport phenomena in the cardiovascular system~\cite{Chahibi2013}. As a result of advection, the virions are transported by the flow of the blood at different velocities in different locations of the cardiovascular system. As a result of diffusion, the virions are transported from a region of higher concentration to a region of lower concentration. This pattern of motion follows the Brownian motion spread in the blood.

To leave the circulatory system and infect other sites in the body, the virions need to penetrate the blood vessel walls made of the \underline{endothelial cells} (Fig.~\ref{fig: ViralSpread}b). We refer to modelling viral translocation across the endothelial barrier as \textsc{\textbf{Model 2}}. Viruses enter and then infect or cross endothelial cells by endocytosis at the apical (luminal) membrane. When infecting the endothelial cell, the virions penetrate the host cell cytosol by fusion from endosomes. However, if the virus crosses the endothelial cell, the virions are transported via intracellular trafficking and ejected from the basolateral (abluminal) membrane into the extracellular space. This leads to the following chemical kinetic model~\cite{Arjmandi2020}:
\begin{eqnarray}
\frac{\partial V^{(b)}(\bar{r},t)}{\partial t} &=& -(c+k_f^1)V^{(b)}(\bar{r},t) + k_b^1V^{(b)}_{\text{BV}}(\bar{r},t) \label{e: mass-equation-first}\\
\frac{\partial V^{(b)}_{\text{BV}}(\bar{r},t)}{\partial t} &=& k_f^1 V^{(b)}(\bar{r},t) - (k_b^1+k_f^2)V^{(b)}_{\text{BV}}(\bar{r},t) + \nonumber \\ &+& k_b^2V^{(b)}_{\text{EV}}(\bar{r},t) \\
\frac{\partial V^{(b)}_{\text{EV}}(\bar{r},t)}{\partial t} &=& k_f^2V^{(b)}_{\text{BV}}(\bar{r},t) - k_b^2V^{(b)}_{\text{EV}}(\bar{r},t) - \nonumber \\ &-& (k_f^3+k_p)V^{(b)}_{\text{EV}}(\bar{r},t) \\
\frac{\partial V^{(b)}_{\bar{\text{V}}}(\bar{r},t)}{\partial t} &=& k_p V^{(b)}_{\text{EV}}(\bar{r},t)\\
\frac{\partial V^{(b)}_{\hat{\text{V}}}(\bar{r},t)}{\partial t} &=& k_f^3 V^{(b)}_{\text{EV}}(\bar{r},t) \label{e: mass-equation-last},
\end{eqnarray}
where $V^{(b)}$, $V^{(b)}_{\text{BV}}$, $V^{(b)}_{\text{EV}}$, $V^{(b)}_{\bar{\text{V}}}$ and $V^{(b)}_{\hat{\text{V}}}$ represent the viral load at the extracellular space (luminal side), the viral load at the endothelial host cell membrane (host cell-bound virions), the viral load in the host cell endosomes, the viral load in the host cell cytosol that penetrated the endosomes, and the viral load at the extracellular space (abluminal side), at point $\bar{r}$ and time $t$, $\bar{r} \in \partial \mathcal{D}$ ($\partial \mathcal{D}$ is the set of points over the endothelial host cell membrane). 
$k_f^i$ and $k_b^i$, $i = 1,2,3$ are forward and backward reaction rates in ms$^{-1}$ and s$^{-1}$, respectively, and $k_p$ is the endosome penetration rate in s$^{-1}$. 

The virion transport mechanism across the blood vessel walls imposes a boundary condition for advection-diffusion in the vessel. This mechanism is modelled by a continuous-time Markov chain framework leading to the following general homogeneous boundary condition~\cite{Arjmandi2020}:
\begin{eqnarray}\label{e:boundary}
    &&D\left(\frac{\partial^2}{\partial t^2}+(k_b^1+k_f^2+k')\frac{\partial}{\partial t} + k_f^2k_f^3 + k_f^2k_p + k_b^1k'\right) \nonumber \\ &\times& \nabla V^{(b)}(\bar{r},t)\cdot\hat{n} \\
     &=& k_f^1\left(\frac{\partial^2}{\partial t^2} + (k_f^2+k')\frac{\partial}{\partial t} + k_f^2k_f^3 + k_f^2k_p \right)V^{(b)}(\bar{r},t), \nonumber
\end{eqnarray}
where $D$ is the diffusion coefficient in m$^2$s$^{-1}$ of the virions in the blood, $k' = k_b^2 + k_f^3 + k_p$, $\nabla$ is the gradient operator, ($\cdot$) is the inner multiplication operator, and $\hat{n}$ is the surface normal at $\bar{r} \in \partial \mathcal{D}$ pointing towards the vessel luminal side. The virion advection-diffusion in the observed blood vessel can then be modelled by the Fick's second law:
\begin{eqnarray}
D\nabla^2 V^{(b)}(\bar{r},t) - c_bV^{(b)}(\bar{r},t) - \bar{v}(\bar{r})\cdot\nabla V^{(b)}(\bar{r},t) \nonumber \\ 
+ S(\bar{r},t) = \frac{\partial V^{(b)}(\bar{r},t)}{\partial t},
\end{eqnarray}
subject to the boundary condition (\ref{e:boundary}). The release rate of the virus at point $\bar{r}$ is given by the source term $S(\bar{r},t)$  (virion s$^{-1}$m$^{-1}$), $\nabla^2$ is the Laplace operator, and $c_b$ is a constant rate of viral clearance in the blood. The blood is assumed to have a laminar flow in the axial direction with uniform velocity profile $\bar{v}(\bar{r}) = v\hat{a_z}$ ms$^{-1}$, where $\hat{a_z}$ is the axial unit vector. The fundamental characteristic function for advection-diffusion called the concentration Green’s function is analytically derived in terms of a convergent infinite series~\cite{Arjmandi2020}. The obtained concentration Green’s function is coupled to the boundary condition given in (\ref{e:boundary}) and provides a useful tool for prediction of the viral load in blood vessels.

Viruses rarely enter into \textbf{neurons} directly to evoke the neurotropic viral spread (Fig.~\ref{fig: ViralSpread}c). We refer to modelling neurotropic viral spread as \textsc{\textbf{Model 3}}. Viruses first replicate locally and then infect nerves associated with the tissue~\cite{Louten2016}. Thus, viruses first infect neurons of the peripheral nervous system, and then gain access to the central nervous system. There is emerging speculations that the central nervous system may be involved during SARS-CoV-2 infection, where neuron-to-neuron transmission route is used to spread the virus~\cite{Huang2020}. Other examples of neuroinvasive viruses include several herpes viruses (e.g., herpes simplex virus) and poliovirus, which is weakly neuroinvasive, and rabies virus, which requires tissue trauma to become neuroinvasive. The literature is, however, very sparse concerning biological models on the neurotropic viral spread. Therefore, we advocate for the consideration of detailed biological models where the characterisation of viral spread throughout the nervous system is considered in more details. This includes addressing secondary mechanisms evolved by some viruses to help them replicate and spread (e.g., binding to a host cell protein called dynein which then transports viral capsid to the neural nucleus for replication).

\textbf{Extracellular vesicles (EVs)} are exchanged between all cells and emerge as the novel, yet obscure cell-to-cell communication mediators. EVs vary in size (50--5000 nm) and contain and transport transmembrane proteins in their lipid bilayer, as well as the cytosol molecular components from the parental cell. The latter includes functional proteins, lipids, and genetic materials [e.g., messenger RNA (mRNA), non-coding RNA (ncRNA), and DNA]~\cite{Kalluri2020}. EVs can also transfer functionally active cargo and have the ability to participate in biological reactions associated with viral dissemination -- the evidence exists for HCV, HIV, and Epstein-Barr virus (EBV) -- and immune response (Fig.~\ref{fig: ViralSpread}d)~\cite{Urbanelli2019}. 

\begin{figure*}[t!]
    \centering
    \includegraphics[width = \linewidth, trim=0cm 7cm 0cm 0cm]{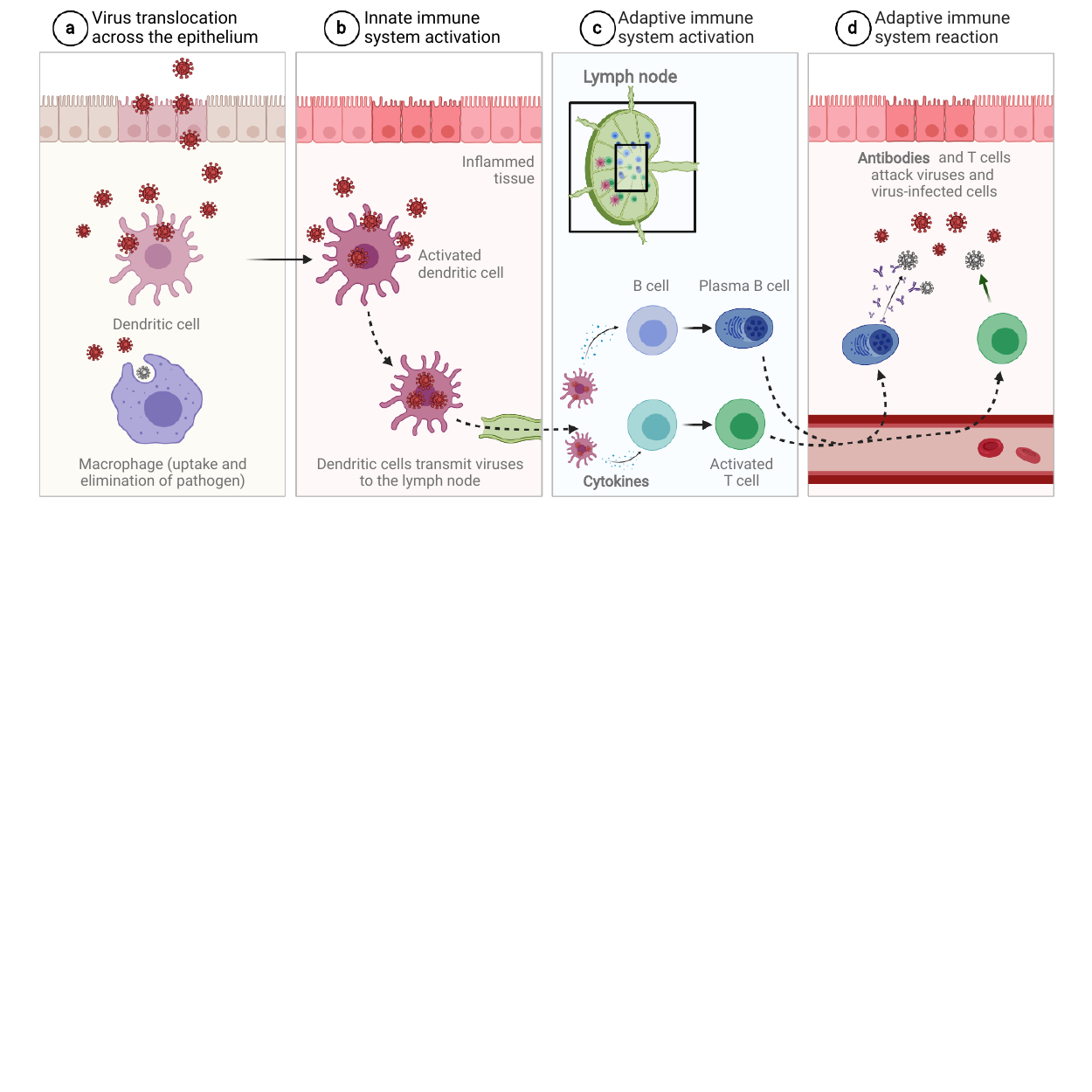}
    \caption{Activation of innate and adaptive immune systems. The cytokine-based- and antibody-based Molecular Communications systems are shown in Phase (c) and (d), respectively.}
    \label{fig: ImmuneSystem}
\end{figure*}

EVs and viruses share common features in their size, structure, biogenesis and uptake~\cite{Hoen2016}. EVs either favour viral infections or limit them, by prompting viral spread or modulating the immune response, respectively. When leveraging viral infection, virus-associated EVs deploy mechanisms such as the delivery of (a) proteins that make the cell more susceptible to infection, (b) viral receptors to cells that are devoid of these receptors thus allowing cells to be infected, (c) nucleic acids that improve and sustain the production of a virus, and (d) molecules that eliminate the host protein relevant for an antiviral response~\cite{Hoen2016, Urbanelli2019}. On the other hand, different mechanisms can be activated by the EVs released by infected cells to prompt an immune response against viruses. The most important mechanisms are the spreading of viral antigens via EVs, and the transfer of cytosolic proteins and nucleic acids involved in antiviral responses. 
Nonetheless, it is still unclear what cell conditions and virus types release EVs that favour or fight infection. 

For initial Molecular Communications system modelling, it seems that viral components hijack the EV secretory routes to exit infected cells and use EV endocytic routes to enter uninfected and immune system cells~\cite{Urbanelli2019}. We refer to modelling EV-based viral spread as \textsc{\textbf{Model 4}}. Each infected cell in this model serves as the transmitter, actively interacting with other cells~\cite{Veletic2020}. The transmitting cell either 1) produces EVs (specifically, exosomes) through its intracellular machinery and releases them upon the fusion of intermediate vesicle-containing endosome compartments, referred to as multivesicular bodies, with the plasma membrane, or 2) involves vertical trafficking of molecular cargo to the plasma membrane, a redistribution of membrane lipids, and the use of contractile machinery at the surface to allow for vesicle pinching (specifically, microvesicles)~\cite{Tricarico2017}. This corresponds to EVs moving from the intracellular space to the extracellular space (the propagation medium). The aspects of EV release yet need to be theoretically investigated addressing infection factors. 

The extracellular matrix (ECM) is the interstitial channel through which EVs are exchanged between the infected virus-producing transmitting cells and the target uninfected receiving cells. The ECM is a 3D molecular network composed of macromolecules. To reach the targeted cell, both the virions and virus-associated EVs should navigate around these macromolecules and diffuse inside and outside other cells in the ECM. The Langevin stochastic differential equation (SDE) can potentially be utilised as a channel modelling tool~\cite{Akyildiz2019}. Since EVs propagate within the ECM based on a drifted random walk, the Langevin SDE needs to contain contributions from the Brownian stochastic force and the drift velocity of the interstitial fluid. Besides, the Langevin SDE needs to be modified to address (a) the losses or clearances of EVs via uptake from other cells and/or degradation through enzymatic attacks, and (b) the anisotropic EV diffusion affected by the ECM properties, i.e., \textit{volume fraction} and \textit{tortuosity}. The volume fraction defines the percentage of the total ECM volume accessible to the virus-bearing EVs. The tortuosity describes the average hindrance of a medium relative to an obstacle-free medium. Hindrance results in an effective diffusion that is decreased compared with the free diffusion coefficient of EVs.

The receiving cell takes up EVs once they bind to the cell-membrane utilising one of the three mechanisms: 1)\textit{ juxtacrine signalling} – where EVs elicit transduction via intracellular signalling pathways, 2) \textit{fusion} – where EVs fuse with the cellular membrane and transfer cargo (i.e., virus-associated components) into the cytoplasm, and 3) \textit{endocytosis} – where EVs internalise and retain in transport vesicles. Non-linear EV-uptake associated with these various mechanisms have been initially investigated in terms of EV-based drug delivery, utilising the Volterra series and multi-dimensional Fourier analysis~\cite{Veletic2019}. The ability to receive viral loads and react accordingly can serve as the performance indicator to reconstruct the information sent by the transmitting cell. 

\subsubsection{Immune System Response}
The immune system is a complex network of cells and proteins that defends the body against pathogen infection. Two subsystems compose the immune system: the \textit{innate immune system} and the \textit{adaptive immune system}. The innate immune system is referred to as non-specific as it provides a general defence against harmful germs and substances. The adaptive immune system is referred to as specific as it makes and uses antibodies to fight certain germs that the body has been previously exposed to. 

The immune system thus works to eradicate the virus. Considering a detailed role of the immune system, i.e., the additional mechanisms in fighting a viral infection $V^{(l)}$ from the \textit{innate immune response} (IIR) and \textit{adaptive immune response} (AIR), ODEs (\ref{e: virusLocal})-(\ref{e: targetLocal}) can be extended~\cite{Handel2010}:
\begin{eqnarray}
\frac{\mathrm{d}V^{(l)}(t)}{\mathrm{d}t}&=& \frac{p}{1+\epsilon_pR_{IIP}(t)}I_2(t) - cV^{(l)}(t) - \nonumber \\
&-& kV^{(l)}(t)T(t) - hV^{(l)}(t)R_{AIR}(t) \\
\frac{\mathrm{d}I_1(t)}{\mathrm{d}t}&=& kV^{(l)}(t)T(t) - \omega I_1(t) \\
\frac{\mathrm{d}I_2(t)}{\mathrm{d}t}&=& \omega I_1(t) - \delta I_2(t) \\
\frac{\mathrm{d}T(t)}{\mathrm{d}t} &=& rD(t) - kV^{(l)}(t)T(t) \\
\frac{\mathrm{d}R_{IIR}(t)}{\mathrm{d}t} &=& \psi V^{(l)}(t) - bR_{IIP}(t) \label{e: IIR} \\
\frac{\mathrm{d}R_{AIR}(t)}{\mathrm{d}t} &=& f V^{(l)}(t) + \beta R_{AIP}(t) \label{e: AIR}.
\end{eqnarray}
Additional effects are also included: two populations of infected cells -- infected but not yet virus-producing cells ($I_1$) with the duration of latent eclipse phase of $1/\omega$, and infected and virus-producing cells ($I_2$), as well as dead cell ($D$) replacement by new susceptible cells at a constant rate $r$. The IIR ($R_{IIR}$) frees the virus at a constant rate $\psi$ and dies at a constant rate $b$; $\epsilon_p$ is the strength of innate response. The AIR ($R_{AIR}$) is activated proportional to the free viral load at a constant rate $f$. Activation is followed by clonal expansion at a constant rate $\beta$. The AIR neutralises the virus with a constant rate $h$.

$R_{IIR}$ and $R_{AIR}$ in (\ref{e: IIR}) and (\ref{e: AIR}) represent concentrations of cytokines and antibodies, respectively. \textbf{Cytokines} are peptides secreted by immune cells (predominantly macrophages, dendritic cell and T-helper cells) \textit{to orchestrate an immune response or an attack on the invading pathogen} (Fig.~\ref{fig: ImmuneSystem}c). Cytokines spread through the body and attach to surface receptors of other immune cells. The receptors then signal the cell to help fight the infection. Cytokines are divided into four categories -- interleukins, interferons, chemokines and tumour necrosis factors -- which can be pro-inflammatory or anti-inflammatory, thus promoting or inhibiting the proliferation and functions of other immune cells. \textbf{Antibodies} are unique proteins encoded by millions of genes which are made and mutated in the human body. They are secreted by immune cells (predominantly plasma B cells differentiated from B cells) \textit{to neutralise the pathogen} (Fig.~\ref{fig: ImmuneSystem}d). The antibody neutralises the pathogen by recognising a unique molecule of the pathogen, called an antigen, via the fragment antigen-binding (Fab) variable region.

Similar to the virus analysis, the Molecular Communications paradigm can give us a clear understanding of how the immune system acts and develops within the body over time. Without going into detailed elaboration, we identify the following two systems:
\begin{itemize}
    \item The \textit{Cytokines-based Molecular Communications system}, which represents cells like macrophages, T-helper cells, natural killer cells, neutrophils, dendritic cells, mast cells, monocytes, B cells and T cells, all serving as transceivers; we refer to modelling cytokines-based Molecular Communications system as \textsc{\textbf{Model 5}}, and
    \item The \textit{Antibody-based Molecular Communications system}, which represents cells like plasma B cells and T cells serving as transmitters, and the virus serving as the receiver; we refer to modelling antibody-based Molecular Communications system as \textsc{\textbf{Model 6}}.
\end{itemize} 
In the context of communications, the cytokines and antibodies are thus considered as information carriers, which propagate messages from the location of transmission until the location of the reception. The information conveyed by the cytokines and antibodies is \textbf{infection reaction}, as a response to infection action. 

\subsection{Out-body Molecular Communications Models} \label{sec:outbody}

The airborne spread of infection is the main mechanism of human-human transmission of viruses. Once viruses are excreted into the air, they propagate towards another person that inhales them into its lungs. This mechanism allows the virus infection spreading to local or pandemic levels, which can occur in a matter of days. There are other modes of human-human transmission of viruses, including human contact transmission, or sexual transmission, but we do not explore these modes of transmission in this paper. Our objective in this section is to explore and analyse the airborne virus Molecular Communications system. It is comprised of the human excretion system as the virus transmitter, the propagation of the virus in the air as the channel, and the human respiratory system as the receiver. 


\subsubsection{Transmitter}

\begin{figure}[t!]
    \centering
    \includegraphics[width = 1\linewidth, trim=0cm 8cm 6cm 0cm]{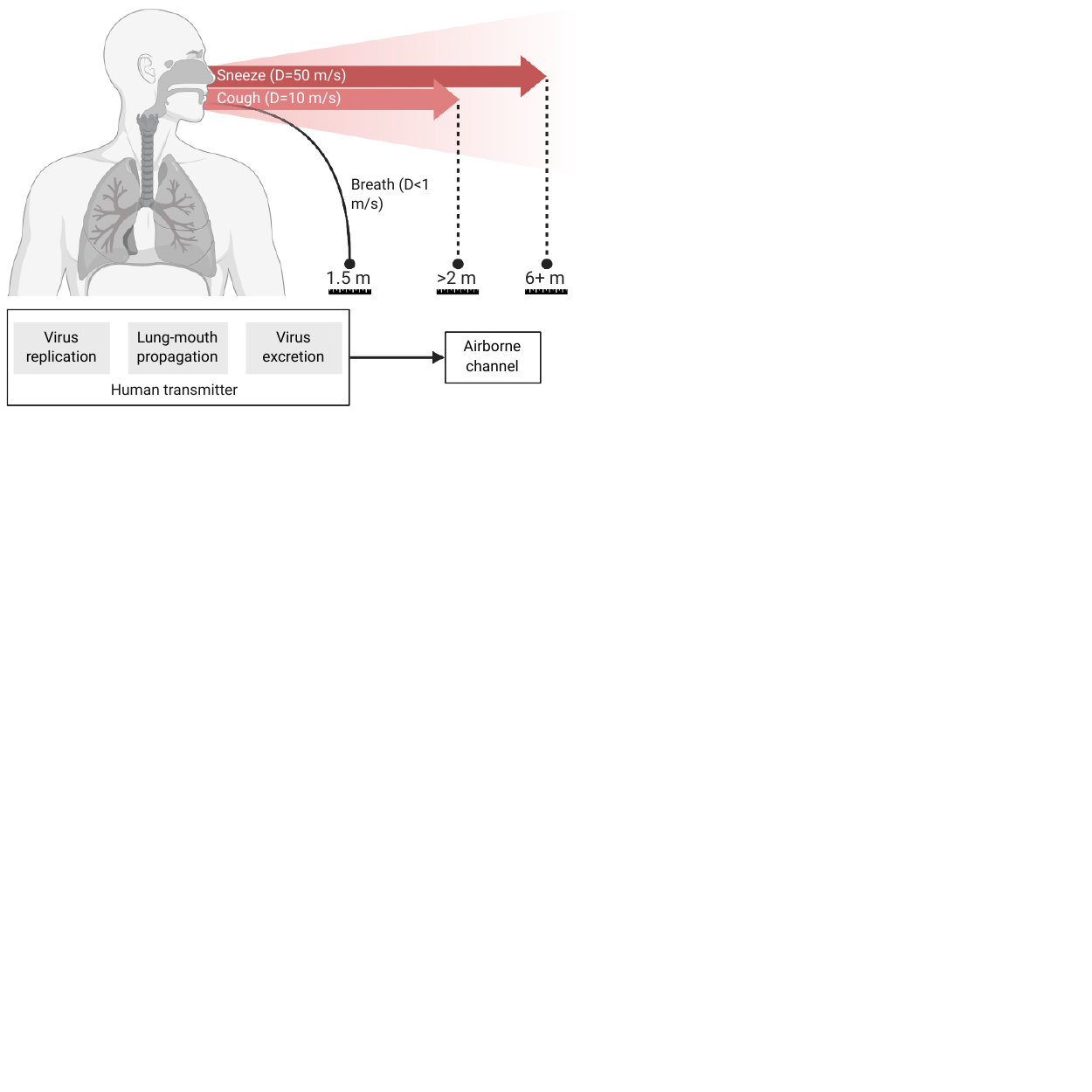}
    \caption{Molecular Communications model of a human transmitter of airborne viruses. The system is comprised of the virus replication, lung-mouth propagation and virus excretion phases. They dictate the rate and strength by which the virus is released to the environment that leads to different range in propagation distance. The sneeze, cough and breath are three different transmission modes for virus excretion.}
    \label{fig:tx}
\end{figure}

We consider the human as the source of virus transmission in the air. The infected humans excrete the virus with a particular concentration rate and velocity via the respiratory system. The respiratory system is composed of the nasal/oral cavity, pharynx, larynx, trachea and lungs, which are comprised of the bronchus and alveolus. Excretion of the virus starts from the alveolus, and propagation to the bronchus towards the nasal/oral cavity. 

Recent works in the literature have been reporting several models that describe the release of particles or droplets in the air by the respiratory system. For example, the model presented in \cite{khalid2020modeling} explores the release of droplets by breath, sneeze and cough. The authors consider a rate model together with an event profiler to condition the rates of droplet release based on the three modes of transmission. The authors are interested in the steady-state derivations where these three modes converge to an averaged exhalation process. We believe that steady-state models do provide attractive mathematical solutions; however, they do not consider the generation process of the droplets and phenomena that influence the fate of the droplets apart from diffusion properties. In a similar goal but with a different approach, another initial transmitter model that analyses the air cloud produced by events driven by exhalation processes was proposed \cite{gulec2020molecular}. It withstands the same issues with the previous model, where the release rates neglect different phenomena that influence the droplets, and in this particular case, the cloud and its characteristics. The work developed in \cite{bourouiba2014violent} showed that the droplets evolve inside a turbulent jet transitioning shortly to a puff. Ejected droplets are surrounded by a dynamically evolving air volume that is coupled to the droplet trajectory. While the major interest has been paid to static or averaged conditions of droplets, we argue that the literature fails to address (a) how the generation of droplets by the human body is coupled with existing modelling efforts and (b) how the conditions of the human body of the infected person impact the exhalation of the droplets in the air. Therefore, we advocate for the consideration of biological models, or variables adjusted from them, where the characterisation of the droplets is thoroughly considered while having a more detail process of how they are generated. 

In this paper, we concentrate on the analysis of three different modes of transmission models for the virus excretion processes, which include breathing, coughing and sneezing, as depicted in Fig.~\ref{fig:tx}. These modes dictate the initial properties of the virus propagation in the air, which alters the outcome of its range of propagation and the velocity that the virus diffuses. In \cite{khalid2019communication}, the authors compiled these modes of virus excretion under one umbrella and referred to them as exhaled breathing. However, more details should be provided as to how these different modes of transmission impact virus propagation. We explain more in the following, where we provide an initial Molecular Communications model of this transmission process which we refer to as \textsc{\textbf{Model 7}}.

We consider that the concentration of droplets $D_i$ released by the human transmitter is modelled as a convection process of virus concentration $V_i$ in the lungs and enters the nasal/oral cavity with rate $k_c$, representing the virus excretion process. We define this process as follows: 
\begin{equation}\label{eq:droplet}
     \frac{\mathrm{d} D_i}{\mathrm{d}t} = \nabla D_dk_c V_i (\nabla D_i),
\end{equation}
\noindent and,
\begin{equation}\label{eq:virus_conc}
    \frac{\mathrm{d} V_i}{\mathrm{d}t} = \nabla V_0D_d(\nabla V_i) + (V_r(t)\circledast P_v(t))
\end{equation}
where $V_0$ is the initial concentration of virus in the nasal/oral cavity, $V_r(t)$, is the rate of virus replication in the lungs, $P_v(t)$ is the propagation of virus from the lungs to the nasal/oral cavity, and $\circledast$ is the convolution operator. We do not explore this model in detail, since the detailed version can be found in \cite{bourouiba2014violent,khalid2020modeling}. However, we acknowledge that $k_c$ is directly linked with the transmission modes that we discussed earlier (breathing, coughing and sneezing). Typically, these modes are always considered to affect the velocity of propagation of droplets. We argue that they are fundamental in the characterisation of virus conversion to droplets as well as the rate of released droplets themselves. Many approaches do consider $k_c$ in the steady-state, but we like to draw attention by the community that it can have non-linear relationships with the infected humans, so it should also be associated with different disease stages over time that is bounded by transition probabilities of stage changes (severe stage to mild, and vice versa). Moreover, as shown in both (\ref{eq:droplet}) and (\ref{eq:virus_conc}), that  non-linearities do exist, for example, the relationship $V_r(t) \circledast P_v(t)$ is added to the model but is currently non-existing in the literature.


\subsubsection{Channel}
Droplets travel in the air following diffusion properties bounded by airflow properties. For example, the modes of transmission discussed above can impact on different types of turbulent flows that lead to a puff scenario, further resulting in airflow forces that are weakened and gravitational forces on the droplet to get stronger. These droplets tend to travel a few meters away from the human excretion point (transmitter), which takes around several seconds, and can reach the human receiving points (e.g., nose or mouth) up to 6 meters of distance. Molecular Communications channel models can be used to model these effects of droplet concentration release in the air, and possibly be used to characterise the number of delivered droplets to the human receiving points. In this section, we review existing models and analyse a general channel model.

\begin{table*}[t!]
\centering
\caption{Literature review summary on channel models for virus air propagation}
\begin{tabular}{c|c|l c c c c}\toprule
\textbf{Propagation mode} & \textbf{Ref.} & \textbf{Medium} & \textbf{Turbulent Flow} & \textbf{Puff Flow} & \textbf{Droplet Evaporation} & \textbf{Droplet Crystallisation} \\ \midrule
Air- & \cite{chaudhuri2020modeling}     &  Transient Air       &      \checkmark          &    \checkmark       &   \checkmark          &   \checkmark              \\
based &\cite{gulec2020molecular} & Air Cloud &                & \checkmark &             &           \\ \midrule
&\cite{chen2010some}& Single Droplet &                & \checkmark  & \checkmark &           \\  
Molecular-&\cite{han2013characterizations} & Particle Dist. & \checkmark & \checkmark & \checkmark &           \\  
based&\cite{mui2009numerical} & Concentration & \checkmark & \checkmark &             &           \\ 
&\cite{khalid2020modeling} & Concentration/Rate & \checkmark & \checkmark &             &           \\\bottomrule  
\end{tabular}\label{tb:channel}
\end{table*}

We summarise the literature on existing droplet propagation models with certain properties of the droplet and airborne viral particles in Table \ref{tb:channel}. Even though they are not entirely classified as pure Molecular Communications channels, they present not only the physical modelling of droplet propagation but also the effects of droplet propagation and thus can be considered by the community for characterising Molecular Communication Systems. We analyse the literature in terms of the completeness of the physics that govern the droplet propagation. First, we look at their modes of propagation, either air-based or molecular-based. These modes dictate the way these models are constructed. Then, we classify each type of medium that can be utilised for modelling these Molecular Communications approaches. The types of species include airflow behaviour with average droplet concentration (transient air, air cloud) to more focused on the characterisation of the number of droplets (single, distribution, concentration and concentration/rate). We also analyse the airflow properties that are mostly secreted from a person as the transmitting point, this includes turbulent flow and puff flow. Turbulent flow accounts for the advection-diffusion of particles that are influenced by a force, in this case, the air turbulence and flow created from the transmitting point. The puff flow can be regarded as the Brownian diffusion in the air and can be influenced by gravity. Lastly, we analyse the properties of the environment that affects the state of droplets once they are excreted and this includes evaporation and crystallisation. Since the majority of the droplets is comprised of water, it is subject to effects from temperature change that can result in evaporation, as well as the quantity of salt in the droplet that lead to its crystallisation. From Table \ref{tb:channel}, we observe that all species comply with the puff and turbulent flow (\cite{chaudhuri2020modeling,han2013characterizations,mui2009numerical,khalid2020modeling}). However, we do recognise that environmental effects on the droplets are not fully explored for Molecular Communications models. The environmental effects have a significant impact on the propagation of the droplets, as it can either impact (a) on the flow behaviour in space, or (b) on the rate of virus reception by the receiving organ (e.g., nose or lung). Besides future investigation in environmental effects on the droplet propagation, there are also needs for further investigation on the effects of jet streams that affect the viral propagation behaviour. 
This includes understanding the aerodynamic airflow within confined and open areas and how this affects the flow of the viral particle propagation.


\begin{figure*}[t!]
    \centering
    \includegraphics[width = 1\linewidth, trim=0cm 7.6cm 1.2cm 0cm]{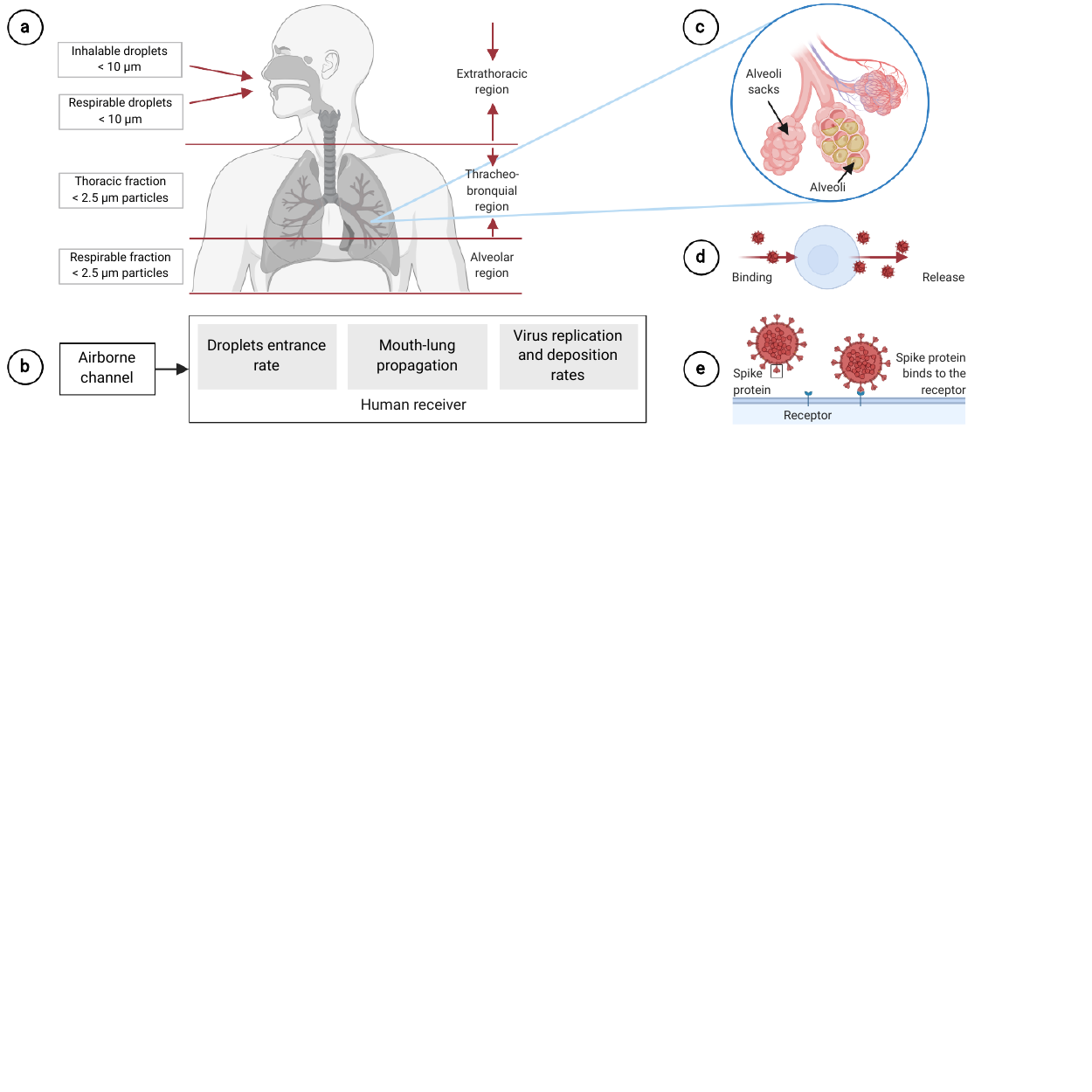}
    \caption{Molecular Communications model of the human receiver of airborne viruses. a) According to the different regions in the respiratory tract, the size of the particles propagating downwards is smaller; b) The human receiver model is comprised of droplets entrance rate, mouth-lung propagation, virus replication and deposition rates; c) Alveoli sack and alveoli with moderate and severe mucus presence due to infection progression; d) The virus duplication process; 2) The virus deposition process.}
    \label{fig:receiver}
\end{figure*}

We now describe the propagation model of airborne droplets, which we refer to as \textsc{\textbf{Model 8}}. 
We assume the source is located at $\vec{r} = [x,y,z]$ and emits droplets with rate $S(\vec{r},t)$. Based on the Fick's second law of diffusion, we consider a droplet concentration varying over time with
\begin{equation} \label{eq:channel}
    \frac{\partial D_i}{\partial t} = \frac{\partial S}{\partial t} - \nabla \vec{F} - \sigma,
\end{equation}
\noindent where $\vec{F}$ is the mass, and $\sigma$ is the droplet degradation loss derived from environmental effects. There are many approaches to derive $\vec{F}$, and this largely depends on the environment. For example, the authors in \cite{khalid2020modeling} focus on expanding the term to include Fick's advection and diffusion effects on the flow. We understand that those terms represent the turbulent flow and puff flow, respectively. On the other hand, \cite{chaudhuri2020modeling} presents a more complex model based on the aerodynamics of the airflow, which precisely addresses the turbulent and jet flows that drive droplet propagation. The authors present the validation of their model using experimental data. 

We show in (\ref{eq:channel}) that $\sigma$ influences directly the propagation of droplets from environmental effects such as evaporation or crystallisation. This is an interesting effect where the temperature, water content, and the salt crystals concentration in the droplet jointly impact diffusion and the rates or concentration when it reaches the receiver. Even though this phenomenon has been explored before \cite{chen2010some,han2013characterizations}, the authors of \cite{chaudhuri2020modeling} investigated this effect, where they derived a model that couples aerodynamic properties and environmental effects on the droplet propagation behaviour. Their model was also validated based on imaging experiments using an ultrasonic levitator to capture transient dynamics of evaporation and precipitation of the evaporating droplet. 
 

One of the main advantages of modelling virus propagation using the Molecular Communications paradigm is the derivation of particle propagation rates, which can be used to study the viral entry mechanism into a human receiver. In \cite{khalid2020modeling}, the authors utilised the solution in (\ref{eq:channel}), by breaking down diffusion components into three dimensions $(x,y,z)$, and algebraically developing the diffusive and mass matrices based on wind flow. To derive a closed-form solution, the authors considered the steady-state response of the breathing process, where they obtain a closed-form Green's function. The obtained closed-form solution for the droplet concentration is represented as 
\begin{equation} \label{eq:chnrates}
    C(\eta,y,z) = \frac{R}{4u\pi \eta} e^{-\frac{y^2}{4\eta}} \left( e^{-\frac{(z-H)^2}{4\eta}} + e^{-\frac{(z+H)^2}{4\eta}} \right)
\end{equation}
where $\eta$ is the turbulence indicator due to wind sources, $H$ is the height of a person's mouth to the ground, and $u$ is the flow velocity in the $x$ dimension.

These models are attractive from the propagation and system theory point of view. However, the authors do not explore practical results in terms of infection. During exposure to an infected human transmitter, several variables dictate the fate of the human receiver, possibly leading to another infection, and this can include the time of exposure as well as the distance between the person emitting the droplets and the human receiver. The authors in \cite{gulec2020molecular} present an analysis on this end-to-end scenario, where the probabilities of infection at the human receiver are evaluated in terms of distance, time of exposure, and coughing angle. As droplets propagation is bounded by the angle of release, it is also important to incorporate spatial analysis on the infection probability, such as the coughing angle.

\subsubsection{Receiver}
In the human receiver, the airborne droplets are absorbed by the nasal-mouth cavity and propagate along the respiratory tract until they reach the alveoli. During this path, the virus undergoes replication, penetrating deep into the epithelial tissues, and then into the circulatory system to infect other organs and systems. In this section, we focus on the high-level modelling and analysis focused on the infection process through the human respiratory tract virus propagation. 

Modelling the human receiver using a systems theory approach can be increasingly complex since there are two main factors that, individually, comprise of several steps. First, there is the entrance of the droplets containing viruses into the human body and residing in the lungs. Secondly, there is the infection process of internal progression. The model for the viral entry can be found in \cite{vimalajeewa2020silico}, which accounts for virus dispersion along the respiratory tract impacted by factors such as the respiratory rate, viral exposure levels (i.e., the quantity of virus that is inhaled), and the virus particle size. The authors developed a computational model to account for the changes in the virus propagation as it enters and propagates in the respiratory tract. For the virus infection process, \cite{vimalajeewa2020silico} also considers the impact of the immune system and how it influences the overall concentration of the virus within the respiratory tract. The model in \cite{gulec2020molecular} focuses on a high-level generic approach based on probability of infection calculated from the probability of reception of droplets containing the virus. Both models are different in terms of biological details or probabilistic infection estimation solutions. We not only support their integration, and creating a scenario where the probability of infection is dependent on increased biology realism, but also for these models to contribute towards modelling the progression of infection and state of infectivity. Understanding the end-to-end propagation to the infectivity process is a crucial contribution for researchers, because this can lead to insights of molecular interactions at the cellular level and the impact of the infection on subsystems of the body, which in turn can lead to precision medicine in clinical decisions for treatments and patient recovery guidelines. 


Our analysis on the viral impact on the human receiver model is depicted in Fig. \ref{fig:receiver}, which we refer to as \textsc{\textbf{Model 9}}. There are three main blocks in the model: the droplets entrance rate, the mouth-lung propagation, and the virus replication and deposition rates. The general model of virus propagation is based on a governing equation based on the Fick's second law for advection-diffusion investigated in \cite{vimalajeewa2020silico}. This model accounts for the concentration of the virus in a particular branch of the respiratory tract $G_i$, and can be represented as
\begin{equation}
    \frac{\partial G_i(x,t)}{\partial t} = D\frac{\partial G_i(x,t)^2}{\partial x^2} - u\frac{\partial G_i(x,t)}{\partial x} - (p-k)G_i(x,t)
\end{equation}
\noindent where $i$ is the generation number ($i=0,1,2,3,...23$) of the lung branches, $k$ is the virus deposition rate, $p$ is the virus replication rate, $D$ is the diffusion coefficient, and $x$ is the direction of the virus propagation (downward or upward in the respiratory tract). The $G_i(x,t)|_{t=0}$ represents the droplet entrance rate. Fig. \ref{fig:receiver}d and Fig. \ref{fig:receiver}e depicts the replication and deposition of the virus, respectively. The authors in \cite{vimalajeewa2020silico} also explored the dynamics change in the 
viral pleomorphic size changes during the propagation. The advection-diffusion component is used to model the breathing process to understand the airflow into the lung. Even though they concentrate on COVID-19, it is clear that their proposed model can be applied to other virus types that propagate through the respiratory tract. For the future, such models may need not only experimental validation, but further integration with infection development itself. As shown in Fig. \ref{fig:receiver}c, the changes in the virus infection process for SARS-CoV-2 (e.g., from moderate to severe), can present changes in the volume of mucus present in the alveolus and the entire respiratory tract. This infection process can not only produce changes in the breathing rhythms and cough rate, but it can also change the advection-diffusion process within the lung generation. As the virus penetrate areas with a high quantity of mucus, the propagation medium changes enough to make the advection property of virus flow to dramatically reduced. Moreover, the deposition and replication rates changes within the mucus areas. This, in turn, can affect the binding process of the virus, which is different for pure diffusion in air pockets compared to a space with mucus. One interesting development could be the investigation of stochastic diffusion coefficients, deposition rates and replication rates of the virus, and how this evolves as the mucus production increases using a multi-medium Molecular Communications model.









\section{Leveraging Experimental Data for Molecular Communications} \label{sec:expdata}



The Molecular Communications scientific community has been abstracting cellular signalling for more than a decade of active research, replicating the characterisation of the functions of wired and wireless networking and computing systems. 
However, with only a few experimental demonstration systems reported to date~\cite{Jamali2019,farsad2013tabletop}, the Molecular Communications field generally falls short on validation. We anticipate similar challenges to arise when building Molecular Communications models of the virus in- and out-body propagation, which are critically needed to understand virus dynamics and unveil new insights that will increase our understanding of virus pathogenesis and enable spread and infection patterns to be more predictable \textit{in vivo}. To characterise the viral dynamics and evolution, one of the initial generic computational modellings demonstrated the necessity to integrate biophysical models and infection properties~\cite{vimalajeewa2020silico}. 

\begin{table*}[t!] 
\centering
\caption{Available quantitative data to support Molecular Communications modelling}
\begin{tabular}{p{1.5cm}|p{7.75cm} p{4.5cm} p{1.75cm} p{0.5cm}}\toprule
\textbf{Type} & \textbf{Viral concentrations in specimens} & \textbf{Characteristics of virion} & \textbf{Detection} & \textbf{Ref.} \\ \midrule 
SARS-CoV (1-2) & \underline{SARS-CoV}: \newline \textbf{Nasopharyngeal   aspirates} -- mean values of 2.3$\times10^5$ RNA copies/mL, 1.9$\times10^7$ copies/mL, and 9.8$\times10^4$ copies/mL on days 5, 10, and 15, respectively (collected from 14 patients) \newline
\textbf{Stool} -- a mean value of 1.0$\times10^7$ copies/mL on day 10 (collected from 1 patient)\newline
\underline{SARS-CoV-2}: \newline 
\textbf{Lower respiratory specimens} (sputum samples) -- 641-1.34$\times10^{11}$ copies/mL, with a median of 7.52$\times10^5$ (collected from 80 patients)\newline 
\textbf{Posterior oropharyngeal saliva} -- a median of 1.58$\times10^5$ copies/mL \newline 
\textbf{Upper respiratory specimens} (nasal and throat swabs) -- cycle threshold (Ct) values 19-40 (collected from 14 patients); Ct values are inversely related to viral RNA copy number, with Ct of 30.76, 27.67, 24.56, and 21.48 corresponding to 1.5$\times10^4$, 1.5$\times10^5$, 1.5$\times10^6$, and 1.5$\times10^7$ copies/mL                                                       &  Enveloped, spherical, 60-140 nm in diameter with spikes of about 9-12 nm \newline \textbf{Genome}: ~30 kb positive-sense, single-stranded RNA \newline
\textbf{RNA Transcript}: 5' cap, 3' poly-A tail \newline
\textbf{Proteome}: 10 proteins & Viral RNA and protein quantification; Serologic detection of antibodies against viruses & \cite{Peiris2003,Hung2004,Pan2020,To2020,Zou2020}, \cite{pickett2012vipr} \\\midrule
   MERS-CoV & \textbf{Lower and upper respiratory specimens} (sputum or tracheal aspirate, and throat-swab) -- a median of 1.62$\times10^7$ RNA copies/mL in the severe group and 3.47$\times10^5$ copies/mL in the mild group (depending on whether oxygen supplementation was used) (P=0.06) (collected from 17 patients) & Around 100 nm in diameter \newline
\textbf{Genome}:  (+)ssrNA, ~30 kb (20-33 kb) enveloped, positive-sense, single-stranded RNA virus \newline
\textbf{Proteome}: 1a, 1b, S, e, M, N, Assorted OrFs \newline
\textbf{RNA Transcript}: 5’-capped & Viral RNA and protein quantification; Serologic detection of antibodies against viruses & \cite{Oh2016}, \cite{pickett2012vipr,fehr2015coronaviruses,al2013middle}  \\\midrule
EBOV & \textbf{Serum} -- up to $10^{10}$ RNA copies/mL in samples from the patients who died; up to $10^8$ within the first 8 days after the onset of symptoms; \newline Viremia classified as low, intermediate, or high according to the viral load ($<$1.73$\times10^4$, 1.73$\times10^4$-8.7$\times10^6$, and $>$ 8.7$\times10^6$) GEQ/mL (genome equivalent per mL), respectively) (collected from 632 patients) &  Filamentous, 970 nm long, 80 nm in diameter, enveloped \newline
\textbf{Genome}: approximately 19 kilobase negative-sense, single-stranded RNA \newline
\textbf{Proteome}: 7 sequentially arranged proteins \newline
\textbf{RNA Transcript}: 5' methyl cap, 3' poly-A tail & Viral RNA and protein quantification; Serologic detection of antibodies against viruses & \cite{Towner2004, deLaVega2015}, \cite{pickett2012vipr}  \\ \midrule
 DENV & \textbf{Serum} -- $10^3$-$10^{11}$ RNA copies/mL (collected from 228 patients, whose data were stratified by infecting dengue serotype (DENV-1, DENV-2, or DENV-3) and by immune status and clinical manifestation (primary infection dengue fever, secondary infection dengue fever, or secondary infection dengue hemorrhagic fever) & Around 50 nm in diameter, icosahedral, enveloped \newline
\textbf{Genome}: 10.7 kilobase positive-sense, single-stranded RNA \newline
\textbf{Proteome}: single polyprotein, co- \& post-translationally cleaved into 11 mature proteins \newline \textbf{RNA Transcript}: 5’ methyl cap, no 3’ poly-A tail & Viral RNA and protein quantification; Serologic detection of antibodies against viruses & \cite{Ben-Shachar2016, Wang2003, Wang2006}, \cite{pickett2012vipr}\\\midrule
 ZIKV & \textbf{Plasma} -- 6.30$\times10^2$ RNA copies/mL \newline \textbf{Urine} -- 1.26$\times10^3$ RNA copies/mL \newline \textbf{Semen} -- 3.98$\times10^8$ RNA copies/mL (collected 2 weeks after symptom onset from a 32-year-old man) and up to 1.58$\times10^9$ RNA copies/mL measured (collected from 12 men in French Guiana with ZIKV infection) \newline 
 \textbf{Plasma}, \textbf{urine}, and \textbf{saliva} -- ZIKV RNA found in the saliva of only the woman; ZIKV remained detectable in their plasma for about 2 weeks, while the urine of the man was consistently positive until day 14 and that of the woman until day 25; The greater virus load in the urine observed in the man (collected from a 69-year-old man and 66-year-old-woman) & Enveloped, spherical, ~50 nm in diameter with an electron dense core of ~30 nm \newline
\textbf{Genome}: ~10.8 kb positive-sense, single-stranded RNA \newline
\textbf{Proteome}: single polyprotein, co- \& post-translationally cleaved into 11 mature proteins \newline
\textbf{RNA Transcript}: 5’ methyl cap, no 3’ poly-A tail & Viral RNA and protein quantification; Serologic detection of antibodies against viruses & \cite{Mansuy2016,deLaval2017,Fourcade2016}, \cite{pickett2012vipr}\\\midrule
 HCV & \textbf{Serum} -- 2$\times10^5$-1.61$\times10^7$ RNA copies/mL (collected from 38 chronically infected patients -- 19 infected by genotype 1 isolates (1a and 1b), 13 by nongenotype 1 isolates (including genotypes 2 a/c, 3a, and 4), and 6 coinfected by genotype 1 and 6 isolates) & Around 50 nm in diameter, icosahedral, enveloped \newline
\textbf{Genome}: 9.6 kilobase positive-sense, single-stranded RNA \newline
\textbf{Proteome}: single polyprotein, co- \& post-translationally cleaved into 10 mature proteins \newline
\textbf{RNA Transcript}: 5’ internal ribosomal entry site (IRES), no 3’ poly-A tail & Viral RNA quantification; Serologic detection of antibodies against HCV & \cite{Lerat1998}, \cite{pickett2012vipr}\\
\bottomrule
\end{tabular}\label{tab: viralLoadsAndChar}
\end{table*}

\begin{table*}[t!]
\centering
\caption{Viral infection impact and immune markers for entry points} 
\begin{tabular}{p{1.5cm}|p{4.5cm} p{6.5cm} p{3cm} p{0.5cm}}\toprule
\textbf{Type} & \textbf{Infection impact} & \textbf{Specific immune markers} & \textbf{Specific antibodies} & \textbf{Ref.} \\ \midrule
SARS-CoV (1-2) & Initiates by the attachment of ACE2 glycoprotein to host receptors & \underline{SARS-CoV-2}: Lower CD4, CD8, NK cell counts, PD-1, Tim-3 on T cells, Phenotype change in monocytes, IP-10, MCP-3, IL-1RA, IL-6, IL-8, IL-10, IL-2R, IL-1$\beta$, IL-4, IL-18, GM-CSF, IL-2 and IFN-$\gamma$ & Neutralising IgG antibodies that target internal N protein or external S glycoprotein (RBD) & \cite{VABRET2020, Tay2020}  \\\midrule
   MERS-CoV & Initiates by the attachment of a glycoprotein to host DPP4 receptor & IFN-$\alpha$, IL-12, IL-17, IFN-$\gamma$ & RBD of S protein & \cite{Faure2014}\\\midrule
EBOV & Initiates by the attachment of glycoprotein to host receptors & IL-1$\beta$ (high), TNF-$\alpha$ (high), IL-6 (high), IL-1RA, sTNF-R (high), SIL-6R, MIP-1$\alpha$, MIP-1$\beta$, Neopterin, SAA, NO2-, Cortisol, IL-10 & Epitopes on the glycoprotein (RBD) & \cite{Baize2002, Levine2019}  \\ \midrule
 DENV & Initiates by glycoprotein E protein binding to DC-SIGN, a C-type lectin & IFN-$\alpha$, IL-6, IL-15, TNF-$\alpha$, IL-1RA, CXCL8, CCL3, CCL4, CCL5, CCL11, M-CSF, G-CSF, ICAM-1, L-Selectin, P-Selectin, CD40, CD40L, Fas, FasL, TRAIL-R2, TRAIL-R3, RANK-L, C3a, IL-2R, TNF-R1 & Envelope glycoprotein (non-neutralising antibodies may cause ADE) & \cite{Fares-Gusmao2019, Guzman2010} \\\midrule
 ZIKV & Initiates by the attachment of the viral glycoprotein E protein to host receptors & CXCL10, ST2/IL-1R4 & IgM binds ZIKV pre-membrane and envelope glycoprotein & \cite{Fares-Gusmao2019, Collins2019} \\\midrule
 HCV & Initiates by E2 protein interacting with cell surface heparan sulfate proteoglycans & CD127+CD8+ T cells, CD127-CD8+ T cells, CD161+CD8+ T cells, PD-1, Regulatory CD4+ and CD8 T cells, IL-10, CXCR3, CXCR6 & IgG binds HCV E1E2 envelope glycoprotein & \cite{Klenerman2012, Hetta2014, Swann2016}\\\bottomrule
\end{tabular}\label{tab: viralImactAndMarkers}
\end{table*}

Biophysical models present real physiological parameters associated with the physical space where the virus propagates through. These parameters are typically available in the literature to be readily used by the Molecular Communications community for \textit{in silico} modelling. The examples include the analysis of entry and spread of SARS-CoV (1-2) and MERS-CoV in the respiratory system, where \textsc{\textbf{Model 1}} and \textsc{\textbf{Model 7-9}} require values for the airflow profile, and diameter and length of each airway generation available from~\cite{BALASHAZY199115}. For the analysis of entry and spread of DENV, ZIKV and HCV in the circulatory system, \textsc{\textbf{Model 2} and \textsc{\textbf{Model 9}}} require values for the blood velocity profile, diameter and length of each vessel generation available from~\cite{Olufsen2000}. 

The basic set of infection properties includes viral exposure levels (in different specimens), virion characteristics, and how viruses interact with cells and the immune system. We summarise relevant data for SARS-CoV (1-2), MERS-CoV, EBOV, DENV, ZIKV and HCV in Table~\ref{tab: viralLoadsAndChar} and Table~\ref{tab: viralImactAndMarkers}. Apart from exposure levels and virion characteristics (including their genome, proteome and RNA transcripts), Table~\ref{tab: viralLoadsAndChar} identifies detection methods, whereas Table~\ref{tab: viralImactAndMarkers} outlines the infection impact and immune markers for each of the considered viruses. This information is necessary for the effects on the receiver communication models, or \textsc{\textbf{Model 9}}, where the interactions between virus-host cells are found.

Based on the available data, at least the concentration profiles of each of these viruses given in Table~\ref{tab: viralLoadsAndChar}, SARS-CoV (1-2), MERS-CoV and ZIKV translocation across the epithelium, or \textsc{\textbf{Model 1}}, and EBOV, DENV, ZIKV and HCV spread via the bloodstream, or \textsc{\textbf{Model 2}}, can be initially analysed. One way of doing such analyses is to associate the available concentration profiles given in Table~\ref{tab: viralLoadsAndChar} with $V_{o}$ given in (\ref{e: mass-action6}) of \textsc{\textbf{Model 1}} and $V^{(b)}$ given in (\ref{e: mass-equation-first}) of \textsc{\textbf{Model 2}}, respectively, and assume initial values for other relevant concentration profiles occurring in (\ref{e: mass-action6})-(\ref{e: mass-action8}) and (\ref{e: mass-equation-first})-(\ref{e: mass-equation-last}). In addition, the parametric values for the virus interaction with the host cells and immune system cells, e.g., forward- and backward reaction rates, are still unavailable and need to be assumed. Obtaining these values from the experiments is challenging but can lead to accurate predictions of viral dynamics. 

More advanced modelling approaches also operate with the host cell surface receptor distribution, host cell distribution, replication and deposition rates, and immune-response-relevant parameters. For example, in the case of SARS-CoV-2, the virions use the ACE2 receptor to bind to and enter host cells (Table \ref{tab: viralImactAndMarkers}), which is important for \textsc{\textbf{Model 7}} and \textsc{\textbf{Model 9}}. The density of ACE2-expressing host cells is modelled in the literature to follow the Gaussian distribution (e.g., $\mathcal{N}(5.83, 0.71)$ (Copies/mL))~\cite{Xu2020}). Spatial distributions additionally complicate the Molecular Communications modelling since ACE2-expressing host cells are not spread evenly, thus creating a heterogeneous concentration distribution across the respiratory system~\cite{vimalajeewa2020silico}. Other parameters relevant for the target-cell model (presented in Section~\ref{sec:inbody}) that facilitate binding by SARS-CoV-2 virus in the respiratory systems can be found in~\cite{Vargas2020, Wolfel2020}. The overview of the infection impact of other considered viruses is also given in Table~\ref{tab: viralImactAndMarkers}.  

In the case of virus proliferation via neurons and EVs, no relevant data for \textsc{\textbf{Model 3}} and \textsc{\textbf{Model 4}} have been collected in the tables. We thus advocate for the need to conduct relevant experiments and back up the corresponding Molecular Communications models that are yet to be developed. As of the immune system reaction via cytokines and antibodies, we list the relevant immune markers and antibodies for each of the considered viruses in Table~\ref{tab: viralImactAndMarkers}. The corresponding concentration levels are typically available in the literature, e.g., cytokine levels in SARS-CoV~\cite{Zhang2004, Cheung2005, Jiang2005}, and can be used to support \textsc{\textbf{Model 5}} and \textsc{\textbf{Model 6}}.

The airborne virus propagation models through droplets based on \textsc{\textbf{Model 8}} need a set of data to characterise diffusion properties when a human transmitter sneezes, coughs or talks. The availability of this data is a critical issue as we found very limited resources that can be used. However, existing testbeds can be used to generate data that can be used by the Molecular Communications researchers. A particular testbed that is appropriate for this scenario is the \emph{Tabletop Molecular Communication} \cite{farsad2013tabletop}, where the authors presented the release of isopropyl alcohol molecules through an electronic-activated spray. These airborne molecules flow towards an alcohol sensor through the airflow produced by a fan. The authors demonstrate the successful encoding, transmission and reception of information encoded using molecules concentration. This testbed can be used to study the effects of droplet propagation, i.e., coupling it with \textsc{\textbf{Model 8}}. The usage of fans can provide modifications to the airflow that drives the propagation of the droplets, and hence air jet streams found in different indoor or outdoor scenarios. This would help estimate parameters for velocity and turbulence measures, such as $\eta$ and $u$ in (\ref{eq:chnrates}). The production of droplets is not clear, especially when trying to emulate events of sneeze, cough and speech. However, we believe the testbed can be extended to include different versions of the electronic-activated spray that can modulate the release rate of molecules, for example. In that way, parameter $S(\vec{r},t)$ in (\ref{eq:channel}) would be associated with a proper virus release rate profile in each emulated event. Lastly, the propagation of the droplets itself can be used to determine other model parameters including the droplet degradation loss derived from environmental effects, which is $\sigma$ in \ref{eq:channel}. The value of $\sigma$ can be broken down into different components as shown in Table \ref{tb:channel}, where the testbed can be used to estimate $\sigma$. 

\section{Open Issues and Challenges}\label{sec:open}

In this section, we explore the open issues and challenges derived from the discussion presented in our survey so far. Our vision is that Molecular Communications aligns itself further with the area of bioengineering so that models of methods can be verified by experiments, working in close collaboration with relevant experts. We explore linking of experimental data to Molecular Communications models, novel in-body viral intervention techniques, emerging technologies for infection theranostics, bridging Molecular Communications and bioinformatics, and novel Molecular Communications models.

\subsection{Linking Experimental Data to Molecular Communications Models}



As discussed, Molecular Communications needs to provide further integration of its models with experimental data, either already available or through novel experiments. The validation of existing models will be of major benefit to the community as it helps calibrate the works already developed, to be directly applicable to the interpretation or prediction of \textit{in vitro} or even \textit{in vivo} phenomena. Without validation, Molecular Communications models are likely prevented from expanding its usage to other areas of potential applications. There are already some interesting works of how Molecular Communications and experimental data can be integrated into different scenarios and for different applications. For example, models with experimental data can be found for calcium signalling \cite{barros2020engineering,nakano2020roles}, bacteria communication \cite{grebenstein2019molecular,liu2017using,martins2018molecular}, neural communication \cite{balasubramaniam2011development}, and macro particle diffusion \cite{farsad2013tabletop,mcguiness2018experimental}. Other works have made great usage of experimental data, such as data extracted from open source databases and integrated into their Molecular Communication models, and examples include works in \cite{felicetti2014modeling,vishveshwara2009intra}.

The reader must consider referring to all the data provided in Section \ref{sec:expdata} and the references provided for data acquisition for new models of infection using the Molecular Communications paradigm. Specifically, transmitter designs would benefit from the process of genetic information encoding with data from Table \ref{tab: viralLoadsAndChar}. Propagation of viruses can be based on the diffusion information in different medium provided in the same table. Lastly,  the receiver design directly benefits from the information provided in both tables. If the receivers are based on electronic technology, the reader is referred to \ref{tab: viralLoadsAndChar}, but if the receivers are actual biological-based devices (e.g., lung cells), the reader is referred to \ref{tab: viralImactAndMarkers}. However, the data provided here are not sufficient to derive all the necessary variables for required non-linear models with complex behaviour, and for that, the reader should consider them as a guiding basis for the correct understanding of the communication parts involved in Molecular Communications for infections diseases.

\subsection{Novel In-body Viral Intervention Techniques}



Even with the existing efforts in designing vaccination and drugs for viral infections treatment, there is still a need for novel interventions requiring robust in-body solutions \cite{saxena2017mesenchymal}. The main reasons are twofold: 1) the effectiveness of vaccines is not always optimal, and 2) there are consequences to tissues and organs during the infection that might require repairing procedures \cite{saxena2017mesenchymal}. Stem cell-based treatment is a novel in-body solution that has been argued as a leading technology for future intervention. This technology is used in different ways, e.g., can be used for drug delivery, and as regenerative therapy (studies have shown that stem cells can modulate the immune system for patients suffering from SARS-CoV-2 infection). Particular stem cells can be derived from many tissues types, including umbilical cord bone marrow, trabecular bone, synovial membrane, and fetal tissues such as lung, pancreas, spleen, liver, etc. \cite{lee2004isolation,sottile2002stem,de2001multipotent}. By interacting with the media through chemical agents, stem cells can eliminate existing pathogenic behaviours and repair the tissue or organ. 

More specifically, {\emph mesenchymal} stem cell-based approaches have been proposed for interventions in many viral infections, including hepatitis \cite{saxena2017mesenchymal,sato2005human}, and SARS-CoV \cite{gupta2020mesenchymal,islam2020perspective}. Mesenchymal stem cells are adherent, Fibroblast-like cells with the ability of self-renewal and differentiation into multiple cell lineages such as Osteoblasts, Chondrocytes, Adipocytes, and Hepatocytes. Looking at specific lung-damaging infections, Mesenchymal stem cells can secrete IL-10, hepatocyte growth factor, Keratinocytes growth factor, and VEGF to alleviate Acute Respiratory Distress Syndrome (ARDS), regenerate and repair lung damage and resist Fibrosis \cite{gupta2020mesenchymal,cruz2019preclinical,zanoni2019role}. This opens new possibilities for the development of novel Molecular Communications solutions that look at communications between the Mesenchymal stem cells signalling and its process of repairing damaged tissue from the communication process. Similar approaches have been proposed in the community, where Exosome Vesicles (EVs) were used to model the interactions between stem cells and Glioblastoma \cite{Veletic2019, Veletic2020}. Even though EVs are also suggested for the treatment of lung damage \cite{gupta2020mesenchymal}, we recognise that different types of signalling information carriers can be used to create a multi-carrier, or multi-molecules communication system, serving as ways to increase the overall treatment capacity.

\subsection{Emerging Technologies for Infection Theranostics}


In the intersection of fields such as bioengineering, material sciences, and medical sciences lies the development of innovative technology that can lead to efficient treatment or diagnosis of infectious diseases, herein referred to as infection theranostics. Molecular Communications can play a role in these emerging technologies by integrating communications properties of transmission, propagation and reception. For example, microfluidic-based organ-on-a-chip devices provide experimental models of transmission, propagation and reception of molecules from cell-cell, tissue-tissue and even organ-organ, with or without other external molecular agents \cite{perestrelo2015microfluidic}. Another example is the airborne viral detection from biosensors that can be either integrated into all-in-one devices \cite{khalid2019communication} or integrated biosensors with proposed emerging infrastructures for 6G, such as the Intelligent Reflecting Surface  \cite{vsiljak2020evolving}. In this case, virus macro-scale propagation and its reception, through either binding-ligand proteins or electrically charged droplets, can include modelling expertise from Molecular Communications researchers to develop such infrastructures. In the following, we dive deeper into this topic using the above-mentioned examples.

Microfluidic-based organ-on-a-chip are alternative experimental models compared to conventional \textit{in vitro} and animal models, since they capture many tissue structures that are found in human organs 
\cite{tang2020human}. The presented literature investigates how organ-on-a-chip can be used to study virus-host interactions, viral therapy-resistance evolution, and development of new antiviral therapeutics, as well as underlying pathogenesis. This can be applied to   different organs of the human body; for example, infection models have been applied to liver chip \cite{ringehan2017viral}, gut chips \cite{villenave2017human,cifuente2011molecular}, neural chips \cite{johnson20163d} and lung chips \cite{benam2016small,si2019discovery}.
Molecular Communications models of microfluidic-based organ-on-a-chip can be used to develop future infection assays to study virus-induced diseases in real-time and at high resolution. Molecular Communications can aid in inferring methods of disease communication and progression by analysing how cellular molecular functions operate in both healthy as well as infectious states. This can be further extended to design novel molecular modulation mechanisms are used either to augment cellular communication or to understand effects from external molecular signals, such as a viral drug. Within the Molecular Communications community, there are a number of existing research on microfluidics modelling and experiments \cite{deng2017microfluidic,hamidovic2019passive}, and this can be extended to utilise organ-on-a-chip devices for viral theranostics. 

Molecular Communications models can be used to design the sensitivity of the binding process for airborne viral detection technology. The key principle is to couple the modelling of air particle flow (as discussed in Section \ref{sec:outbody}) with the receiver design, which in this case is an airborne viral biosensor. These biosensors can be built in a number of ways, including ligand-binding protein receptors \cite{vsiljak2020evolving}, electrically charged particles  \cite{khalid2019communication}, CMOS-coupled immunological assays, and even through Polymerase Chain Reaction (PCR) technology  \cite{fronczek2015biosensors}. The receiver design for biosensor technology based on a realistic propagation model of viral particles is missing. For example, in an outdoor space, the propagation of the airborne viral particles can undergo stochastic propagation patterns due to the changes in airflow directions and turbulence. Therefore, the design of the biosensor receptor structure can incorporate Molecular Communications model to help enhance the design of the sensors that is appropriate for the specific environment. However, this will require considerable
experimental work where molecular propagation flow can be studied with fluorescent technology, similar to the works proposed in \cite{chaudhuri2020modeling,grebenstein2018biological,mcguiness2018parameter}, in order to characterise the propagation patterns. 

\subsection{Bridging Molecular Communications and Bioinformatics Tools}

In this paper, we mostly investigated the Molecular Communication models of viral infection, but in Section \ref{sec:expdata}, we explored the link between these models to currently available experimental data for viral-host genetic interactions. There exist many interesting works that cover the analysis of these interactions coming from a bioinformatics perspective, which can analyse a large number of protein interactions and how they affect cells activity and fate \cite{roumpeka2017review,soria2015overview,garrido2008evaluation}. 

Bioinformatics is a very advanced area in the analysis of genomics, as previously noted, with several tools and data available for researchers who desire to investigate the genetic relationships of viruses and host cells. Based on the genomic sequencing, these tools can provide the assembly of fragmented sequenced data, phylogenetic analysis of taxonomic groups, identification of genetic structures, identification of domains, functions and metabolic pathways, as well as data sharing capabilities \cite{roumpeka2017review}. Researchers in the area are hoping that, by having open-shared data and tools access policy, they facilitate the discovery of targeted genes and what leads to cell behaviour \cite{wang2015xander}. Some works focus on the discovery of new drugs or even vaccines \cite{soria2015overview,garrido2008evaluation}. Even though some exciting works related to the topic of genomics are presented in the Molecular Communications field \cite{walsh2013reliability,balasubramaniam2013multi,bilgin2018dna}, the total and practical integration of Molecular Communications and bioinformatics tools are far from being explored. More works linking genomics with cell behaviour can solve existing issues, including the characterisation of natural cell signalling modulation techniques, sources of noise and interference, encoding/decoding of molecular information, and synthetic Molecular Communications. The bioinformatics existing tools already provide the information on the cell genetics and, sometimes, linkage to cell behaviour. Seamless integration with Molecular Communications leads to studies such as on the interaction of viral DNA content \cite{walsh2013reliability} and techniques for linking DNA exchange between bacteria for bacteria-based Molecular Communications \cite{balasubramaniam2013multi}. Looking at the various models and systems presented in this paper, we advocate that the viral genetic interactions with host cells can provide novel understandings on how infected cells propagate the genetic content to other cells from an information and communication theory approach, while at the same time also considering the evolution of the genetic content, which can provide novel mechanisms for inferring infection propagation inside the body.

\subsection{Novel Molecular Communications Models}
New advancements in Molecular Communications modelling are needed, from this point in time forward, to help understand the virus propagation both in-body and out-body and the end-to-end communication system. For in-body, the link between virus replication and tissue response has yet to be thoroughly investigated. For example, the interactions of the virus with neural communications \cite{veletic2019information,veletic2019synaptic}, or calcium signalling in different tissues or organs (especially the epithelium) \cite{barros2015comparative,barros2017ca2+}, can provide further understanding between the virus and hosts interactions using molecular signals as infection information carriers. As we know from the literature, the human immune response is triggered after the human body recognises the presence of infectious agents, foreign to the body itself \cite{sompayrac2019immune}. The immune response is yet to be explored using concepts from Molecular Communications, similar to the models initially discussed in Section \ref{sec:background}, and to tie this to other Molecular Communication models (e.g., propagation within the circulatory system), in order to create an end-to-end system model. The key benefit of these studies is the ability to accurately capture the effects of viral propagation communication on the immune response communication, which would create computational models that can benefit vaccine designers in the future if the various sub-communication systems and interactions are well understood. These new models could be used to analyse new techniques to modulate the immune system response, coming from a regenerative medical intervention, based on the precise calculation of infections stages derived from virus-host interactions, in order to lead to an end-to-end diagnostic and therapeutic strategy for vaccine development.

For out-body models, where models have recently been introduced on airborne droplet propagation, there are further developments required. First, transmitter models for out-body Molecular Communications should consider in more detail the conditions of the human transmitter, i.e., the level of infection and condition of the respiratory system. This understanding can provide new relationships between the virus release rates from the host, and the analysis of viral reception concentration by another host. In the receiver, as explored in Section \ref{sec:inbody}, the need to couple models of the propagation of the respiratory tract, with actual rates derived from the airborne droplet propagation models are needed. This will contribute towards and end-to-end model that considers the Molecular Communication airflow propagation coupled with environmental effects on the droplets, and to link this with the in-body propagation of the virus into the lungs. Even though an end-to-end analysis of these systems can lead to increasingly complex models, they can be useful in accurately predicting infection spreading patterns, and how this can impact on people with different health conditions, in order to personalise and classify their risk levels. Such accurate modelling can prevent total lockdowns for the entire society, where people in different categories can be allowed into society provided they take certain protective measures. 

\section{Conclusions}\label{sec:conclusions}
Molecular Communications can play a significant role in viral infectious disease research, by considering the detail characterisation of the transmission, reception and propagation of viruses inside and outside the human body. We provided an extensive review of the existing literature on the topic, by analysing the existing models for in-body and out-body Molecular Communications. For in-body models, we explored the viral translocation across the epithelial and endothelial barriers, neurotropic viral spread, EV-based viral spread, cytokine-based- and antibody-based Molecular Communications. In out-body models, we analysed models for the transmission process of viruses spelt from a human transmitter, the airborne droplet and virus propagation coupled with many environmental effects including turbulent flow, puff flow, droplet evaporation and droplet crystallisation, and finally, the reception process of viruses in the human receiver. Besides, we reviewed models for the virus entrance mechanism focusing on the virus concentration in the respiratory tract. 
We showed how the available experimental data can be integrated into Molecular Communications models, and what are the open issues and future directions. We are looking forward to exciting new research that can come as an output of the interdisciplinary works using Molecular Communications for developing new methods for treatment of infection as well as vaccination methods. Based on the analysis provided in the paper, we are confident that fantastic novel research can emerge and help in the fight against the current and future pandemics.

\bibliographystyle{IEEEtran}
\bibliography{ref}

\begin{thebibliography}{100}
\providecommand{\url}[1]{#1}
\csname url@samestyle\endcsname
\providecommand{\newblock}{\relax}
\providecommand{\bibinfo}[2]{#2}
\providecommand{\BIBentrySTDinterwordspacing}{\spaceskip=0pt\relax}
\providecommand{\BIBentryALTinterwordstretchfactor}{4}
\providecommand{\BIBentryALTinterwordspacing}{\spaceskip=\fontdimen2\font plus
\BIBentryALTinterwordstretchfactor\fontdimen3\font minus
  \fontdimen4\font\relax}
\providecommand{\BIBforeignlanguage}[2]{{%
\expandafter\ifx\csname l@#1\endcsname\relax
\typeout{** WARNING: IEEEtran.bst: No hyphenation pattern has been}%
\typeout{** loaded for the language `#1'. Using the pattern for}%
\typeout{** the default language instead.}%
\else
\language=\csname l@#1\endcsname
\fi
#2}}
\providecommand{\BIBdecl}{\relax}
\BIBdecl

\bibitem{jones2020coronavirus}
L.~Jones, D.~Brown, and D.~Palumbo, ``Coronavirus: A visual guide to the
  economic impact,'' \emph{BBC News}, vol.~28, 2020.

\bibitem{Matina2020coronavirus}
M.~Stevis-Gridneff, ``A €750 billion virus recovery plan thrusts {Europe}
  into a new frontier,'' \emph{The New York Times}, 2020.

\bibitem{saxena2017mesenchymal}
V.~Saxena, ``Mesenchymal stem cell based therapeutic intervention against viral
  hepatitis: A perspective,'' \emph{J Stem Cells and Genetics 2017}, vol.~1,
  no.~1, pp. 1--2, 2017.

\bibitem{thompson1852annals}
T.~Thompson, \emph{Annals of influenza or epidemic catarrhal fever in Great
  Britain from 1510 to 1837}.\hskip 1em plus 0.5em minus 0.4em\relax Sydenham
  society, 1852, vol.~21.

\bibitem{roldao2010virus}
A.~Rold{\~a}o, M.~C.~M. Mellado, L.~R. Castilho, M.~J. Carrondo, and P.~M.
  Alves, ``Virus-like particles in vaccine development,'' \emph{Expert review
  of vaccines}, vol.~9, no.~10, pp. 1149--1176, 2010.

\bibitem{Akyildiz2019}
I.~F. {Akyildiz}, M.~{Pierobon}, and S.~{Balasubramaniam}, ``An information
  theoretic framework to analyze molecular communication systems based on
  statistical mechanics,'' \emph{Proceedings of the IEEE}, vol. 107, no.~7, pp.
  1230--1255, 2019.

\bibitem{nakano2017molecular}
T.~Nakano, ``Molecular communication: A 10 year retrospective,'' \emph{IEEE
  Transactions on Molecular, Biological and Multi-Scale Communications},
  vol.~3, no.~2, pp. 71--78, 2017.

\bibitem{akyildiz2015internet}
I.~F. Akyildiz, M.~Pierobon, S.~Balasubramaniam, and Y.~Koucheryavy, ``The
  internet of bio-nano things,'' \emph{IEEE Communications Magazine}, vol.~53,
  no.~3, pp. 32--40, 2015.

\bibitem{akyildiz2008nanonetworks}
I.~F. Akyildiz, F.~Brunetti, and C.~Bl{\'a}zquez, ``Nanonetworks: A new
  communication paradigm,'' \emph{Computer Networks}, vol.~52, no.~12, pp.
  2260--2279, 2008.

\bibitem{felicetti2016applications}
L.~Felicetti, M.~Femminella, G.~Reali, and P.~Li{\`o}, ``Applications of
  molecular communications to medicine: A survey,'' \emph{Nano Communication
  Networks}, vol.~7, pp. 27--45, 2016.

\bibitem{atakan2012body}
B.~Atakan, O.~B. Akan, and S.~Balasubramaniam, ``Body area nanonetworks with
  molecular communications in nanomedicine,'' \emph{IEEE Communications
  Magazine}, vol.~50, no.~1, pp. 28--34, 2012.

\bibitem{barros2018multi}
M.~T. Barros, W.~Silva, and C.~D.~M. Regis, ``The multi-scale impact of the
  {Alzheimer’s} disease on the topology diversity of astrocytes molecular
  communications nanonetworks,'' \emph{IEEE Access}, vol.~6, pp.
  78\,904--78\,917, 2018.

\bibitem{veletic2019synaptic}
M.~Veleti{\'c} and I.~Balasingham, ``Synaptic communication engineering for
  future cognitive brain--machine interfaces,'' \emph{Proceedings of the IEEE},
  vol. 107, no.~7, pp. 1425--1441, 2019.

\bibitem{balasubramaniam2012realizing}
S.~Balasubramaniam and J.~Kangasharju, ``Realizing the internet of nano things:
  challenges, solutions, and applications,'' \emph{Computer}, vol.~46, no.~2,
  pp. 62--68, 2012.

\bibitem{martins2018molecular}
D.~P. Martins, K.~Leetanasaksakul, M.~T. Barros, A.~Thamchaipenet, W.~Donnelly,
  and S.~Balasubramaniam, ``Molecular communications pulse-based jamming model
  for bacterial biofilm suppression,'' \emph{IEEE Transactions on
  NanoBioscience}, vol.~17, no.~4, pp. 533--542, 2018.

\bibitem{martins2016using}
D.~P. Martins, M.~T. Barros, and S.~Balasubramaniam, ``Using competing
  bacterial communication to disassemble biofilms,'' in \emph{Proceedings of
  the 3rd ACM International Conference on Nanoscale Computing and
  Communication}, 2016, pp. 1--6.

\bibitem{walsh2013reliability}
F.~Walsh and S.~Balasubramaniam, ``Reliability and delay analysis of multihop
  virus-based nanonetworks,'' \emph{IEEE Transactions on Nanotechnology},
  vol.~12, no.~5, pp. 674--684, 2013.

\bibitem{khalid2019communication}
M.~Khalid, O.~Amin, S.~Ahmed, B.~Shihada, and M.-S. Alouini, ``Communication
  through breath: Aerosol transmission,'' \emph{IEEE Communications Magazine},
  vol.~57, no.~2, pp. 33--39, 2019.

\bibitem{khalid2020modeling}
------, ``Modeling of viral aerosol transmission and detection,'' \emph{IEEE
  Transactions on Communications}, 2020.

\bibitem{vimalajeewa2020silico}
D.~Vimalajeewa, S.~Balasubramaniam, D.~P. Berry, and G.~Barry, ``In silico
  modeling of virus particle propagation and infectivity along the respiratory
  tract: A case study for {SARS-CoV-2},'' \emph{bioRxiv}, 2020.

\bibitem{martins2018computational}
D.~P. Martins, M.~T. Barros, M.~Pierobon, M.~Kandhavelu, S.~Balasubramaniam
  \emph{et~al.}, ``Computational models for trapping {Ebola} virus using
  engineered bacteria,'' \emph{IEEE/ACM Transactions on Computational Biology
  and Bioinformatics}, vol.~15, no.~6, pp. 2017--2027, 2018.

\bibitem{Sariol20}
A.~Sariol and S.~Perlman, ``Lessons for {COVID-19} immunity from other
  coronavirus infections,'' \emph{Immunity}, 2020.

\bibitem{florindo2020immune}
H.~F. Florindo, R.~Kleiner, D.~Vaskovich-Koubi, R.~C. Ac{\'u}rcio, B.~Carreira,
  E.~Yeini, G.~Tiram, Y.~Liubomirski, and R.~Satchi-Fainaro, ``Immune-mediated
  approaches against {COVID-19},'' \emph{Nature nanotechnology}, pp. 1--16,
  2020.

\bibitem{lebeau2020deciphering}
G.~Lebeau, D.~Vagner, {\'E}.~Frumence, F.~Ah-Pine, X.~Guillot,
  E.~Nob{\'e}court, L.~Raffray, and P.~Gasque, ``Deciphering {SARS-CoV-2}
  virologic and immunologic features,'' \emph{International Journal of
  Molecular Sciences}, vol.~21, no.~16, p. 5932, 2020.

\bibitem{wadman2020rampage}
M.~Wadman, J.~Couzin-Frankel, J.~Kaiser, and C.~Matacic, ``A rampage through
  the body,'' \emph{Science}, vol. 368, no. 6489, pp. 356--360, 2020.

\bibitem{Ansari14}
A.~A. Ansari, ``Clinical features and pathobiology of {Ebolavirus} infection,''
  \emph{Journal of autoimmunity}, vol.~55, pp. 1--9, 2014.

\bibitem{us2019first}
U.~Food, D.~Administration \emph{et~al.}, ``First {FDA}-approved vaccine for
  the prevention of {Ebola} virus disease, marking a critical milestone in
  public health preparedness and response,'' \emph{Press Announc}, 2019.

\bibitem{henao2017efficacy}
A.~M. Henao-Restrepo, A.~Camacho, I.~M. Longini, C.~H. Watson, W.~J. Edmunds,
  M.~Egger, M.~W. Carroll, N.~E. Dean, I.~Diatta, M.~Doumbia \emph{et~al.},
  ``Efficacy and effectiveness of an {rVSV}-vectored vaccine in preventing
  {Ebola} virus disease: final results from the guinea ring vaccination,
  open-label, cluster-randomised trial ({Ebola {\c{C}}a Suffit!)},'' \emph{The
  Lancet}, vol. 389, no. 10068, pp. 505--518, 2017.

\bibitem{normile2013surprising}
D.~Normile, ``Surprising new dengue virus throws a spanner in disease control
  efforts,'' \emph{Science}, 2013.

\bibitem{conway2016aedes}
M.~J. Conway, B.~Londono-Renteria, A.~Troupin, A.~M. Watson, W.~B. Klimstra,
  E.~Fikrig, and T.~M. Colpitts, ``Aedes aegypti {D7} saliva protein inhibits
  dengue virus infection,'' \emph{PLoS neglected tropical diseases}, vol.~10,
  no.~9, p. e0004941, 2016.

\bibitem{yu2010new}
C.~I. Yu and B.-L. Chiang, ``A new insight into hepatitis {C} vaccine
  development,'' \emph{Journal of Biomedicine and Biotechnology}, vol. 2010,
  2010.

\bibitem{Louten2016}
J.~Louten, ``Virus transmission and epidemiology,'' \emph{Essential Human
  Virology}, pp. 71--92, 2016.

\bibitem{Bomsel2003}
M.~Bomsel and A.~Alfsen, ``Entry of viruses through the epithelial barrier:
  pathogenic trickery,'' \emph{Nature Reviews Molecular Cell Biology}, vol.~4,
  no.~1, pp. 57--68, 2003.

\bibitem{Cong2014}
Y.~Cong and X.~Ren, ``Coronavirus entry and release in polarized epithelial
  cells: a review,'' \emph{Reviews in medical virology}, vol.~24, no.~5, pp.
  308--315, 2014.

\bibitem{Veletic2019}
M.~Veleti\'{c}, M.~T. Barros, I.~Balasingham, and S.~Balasubramaniam, ``A
  molecular communication model of exosome-mediated brain drug delivery,'' in
  \emph{Proceedings of the Sixth Annual ACM International Conference on
  Nanoscale Computing and Communication}, ser. NANOCOM '19.\hskip 1em plus
  0.5em minus 0.4em\relax New York, NY, USA: Association for Computing
  Machinery, 2019.

\bibitem{Zitzmann2018}
C.~Zitzmann and L.~Kaderali, ``Mathematical analysis of viral replication
  dynamics and antiviral treatment strategies: From basic models to age-based
  multi-scale modeling,'' \emph{Frontiers in Microbiology}, vol.~9, p. 1546,
  2018.

\bibitem{Bonhoeffer1997}
S.~Bonhoeffer, R.~M. May, G.~M. Shaw, and M.~A. Nowak, ``Virus dynamics and
  drug therapy,'' \emph{Proc Natl Acad Sci U S A}, vol.~94, no.~13, pp.
  6971--6976, 1997.

\bibitem{Chahibi2013}
Y.~{Chahibi}, M.~{Pierobon}, S.~O. {Song}, and I.~F. {Akyildiz}, ``A molecular
  communication system model for particulate drug delivery systems,''
  \emph{IEEE Transactions on Biomedical Engineering}, vol.~60, no.~12, pp.
  3468--3483, 2013.

\bibitem{Arjmandi2020}
Y.~{Arjmandi}, M.~{Zoofaghar}, S.~V. {Rouzegar}, M.~{Veleti\'c}, and
  I.~{Balasingham}, ``On mathematical analysis of active drug transport coupled
  with flow-induced diffusion in blood vessels,'' \emph{IEEE Transactions on
  NanoBioscience}, vol.~x, no.~x, pp. x--x, 2020, in press.

\bibitem{Huang2020}
J.~Huang, M.~Zheng, X.~Tang, Y.~Chen, A.~Tong, and L.~Zhou, ``Potential of
  {SARS-CoV-2} to cause {CNS} infection: Biologic fundamental and clinical
  experience,'' \emph{Frontiers in Neurology}, vol.~11, p. 659, 2020.

\bibitem{Kalluri2020}
R.~Kalluri and V.~S. LeBleu, ``The biology, function, and biomedical
  applications of exosomes,'' \emph{Science}, vol. 367, no. 6478, 2020.

\bibitem{Urbanelli2019}
L.~Urbanelli, S.~Buratta, B.~Tancini, K.~Sagini, F.~Delo, S.~Porcellati, and
  C.~Emiliani, ``The role of extracellular vesicles in viral infection and
  transmission,'' \emph{Vaccines}, vol.~7, no.~3, p. 102, 2019.

\bibitem{Hoen2016}
E.~Nolte-'t Hoen, T.~Cremer, R.~C. Gallo, and L.~B. Margolis, ``Extracellular
  vesicles and viruses: Are they close relatives?'' \emph{Proceedings of the
  National Academy of Sciences of the United States of America}, vol. 113,
  no.~33, pp. 9155--9161, 2016.

\bibitem{Veletic2020}
M.~{Veleti\'{c}}, M.~T. {Barros}, H.~{Arjmandi}, S.~{Balasubramaniam}, and
  I.~{Balasingham}, ``Modeling of modulated exosome release from differentiated
  induced neural stem cells for targeted drug delivery,'' \emph{IEEE
  Transactions on NanoBioscience}, vol.~19, no.~3, pp. 357--367, 2020.

\bibitem{Tricarico2017}
C.~Tricarico, J.~Clancy, and C.~D'Souza-Schorey, ``Biology and biogenesis of
  shed microvesicles,'' \emph{Small GTPases}, vol.~8, no.~4, pp. 220--232,
  2017.

\bibitem{Handel2010}
A.~Handel, I.~M. Longini, and R.~Antia, ``Towards a quantitative understanding
  of the within-host dynamics of influenza {A} infections,'' \emph{Journal of
  The Royal Society Interface}, vol.~7, no.~42, pp. 35--47, 2010.

\bibitem{gulec2020molecular}
F.~Gulec and B.~Atakan, ``A molecular communication perspective on airborne
  pathogen transmission and reception via droplets generated by coughing and
  sneezing,'' \emph{arXiv preprint arXiv:2007.07598}, 2020.

\bibitem{bourouiba2014violent}
L.~Bourouiba, E.~Dehandschoewercker, and J.~W. Bush, ``Violent expiratory
  events: on coughing and sneezing,'' \emph{Journal of Fluid Mechanics}, vol.
  745, pp. 537--563, 2014.

\bibitem{chaudhuri2020modeling}
S.~Chaudhuri, S.~Basu, P.~Kabi, V.~R. Unni, and A.~Saha, ``Modeling the role of
  respiratory droplets in {COVID-19} type pandemics,'' \emph{Physics of
  Fluids}, vol.~32, no.~6, p. 063309, 2020.

\bibitem{chen2010some}
C.~Chen and B.~Zhao, ``Some questions on dispersion of human exhaled droplets
  in ventilation room: answers from numerical investigation,'' \emph{Indoor
  Air}, vol.~20, no.~2, pp. 95--111, 2010.

\bibitem{han2013characterizations}
Z.~Han, W.~Weng, and Q.~Huang, ``Characterizations of particle size
  distribution of the droplets exhaled by sneeze,'' \emph{Journal of the Royal
  Society Interface}, vol.~10, no.~88, p. 20130560, 2013.

\bibitem{mui2009numerical}
K.~Mui, L.~Wong, C.~Wu, and A.~C. Lai, ``Numerical modeling of exhaled droplet
  nuclei dispersion and mixing in indoor environments,'' \emph{Journal of
  hazardous materials}, vol. 167, no. 1-3, pp. 736--744, 2009.

\bibitem{Jamali2019}
V.~{Jamali}, A.~{Ahmadzadeh}, W.~{Wicke}, A.~{Noel}, and R.~{Schober},
  ``Channel modeling for diffusive molecular communication—a tutorial
  review,'' \emph{Proceedings of the IEEE}, vol. 107, no.~7, pp. 1256--1301,
  2019.

\bibitem{farsad2013tabletop}
N.~Farsad, W.~Guo, and A.~W. Eckford, ``Tabletop molecular communication: Text
  messages through chemical signals,'' \emph{PloS one}, vol.~8, no.~12, p.
  e82935, 2013.

\bibitem{Peiris2003}
J.~Peiris, C.~Chu, V.~Cheng, K.~Chan, I.~Hung, L.~Poon, K.~Law, B.~Tang,
  T.~Hon, C.~Chan \emph{et~al.}, ``Clinical progression and viral load in a
  community outbreak of coronavirus-associated {SARS} pneumonia: a prospective
  study,'' \emph{The Lancet}, vol. 361, no. 9371, pp. 1767 -- 1772, 2003.

\bibitem{Hung2004}
I.~F.~N. Hung, V.~C.~C. Cheng, A.~K.~L. Wu, B.~S.~F. Tang, K.~H. Chan, C.~M.
  Chu, M.~M.~L. Wong, W.~T. Hui, L.~L.~M. Poon, D.~M.~W. Tse \emph{et~al.},
  ``Viral loads in clinical specimens and {SARS} manifestations,''
  \emph{Emerging infectious diseases}, vol.~10, no.~9, pp. 1550--1557, 2004.

\bibitem{Pan2020}
Y.~Pan, D.~Zhang, P.~Yang, L.~L.~M. Poon, and Q.~Wang, ``Viral load of
  {SARS}-{C}o{V}-2 in clinical samples,'' \emph{The Lancet Infectious
  Diseases}, vol.~20, no.~4, pp. 411--412, 2020.

\bibitem{To2020}
K.~K.-W. To, O.~T.-Y. Tsang, W.-S. Leung, A.~R. Tam, T.-C. Wu, D.~C. Lung,
  C.~C.-Y. Yip, J.-P. Cai, J.~M.-C. Chan, T.~S.-H. Chik \emph{et~al.},
  ``Temporal profiles of viral load in posterior oropharyngeal saliva samples
  and serum antibody responses during infection by {SARS}-{C}o{V}-2: an
  observational cohort study,'' \emph{The Lancet Infectious Diseases}, vol.~20,
  no.~5, pp. 565--574, 2020.

\bibitem{Zou2020}
L.~Zou, F.~Ruan, M.~Huang, L.~Liang, H.~Huang, Z.~Hong, J.~Yu, M.~Kang,
  Y.~Song, J.~Xia \emph{et~al.}, ``{SARS}-{CoV}-2 viral load in upper
  respiratory specimens of infected patients,'' \emph{New England Journal of
  Medicine}, vol. 382, no.~12, pp. 1177--1179, 2020.

\bibitem{pickett2012vipr}
B.~E. Pickett, E.~L. Sadat, Y.~Zhang, J.~M. Noronha, R.~B. Squires, V.~Hunt,
  M.~Liu, S.~Kumar, S.~Zaremba, Z.~Gu \emph{et~al.}, ``{ViPR}: an open
  bioinformatics database and analysis resource for virology research,''
  \emph{Nucleic acids research}, vol.~40, no.~D1, pp. D593--D598, 2012.

\bibitem{Oh2016}
M.-d. Oh, W.~B. Park, P.~G. Choe, S.-J. Choi, J.-I. Kim, J.~Chae, S.~S. Park,
  E.-C. Kim, H.~S. Oh, E.~J. Kim \emph{et~al.}, ``Viral load kinetics of {MERS}
  coronavirus infection,'' \emph{New England Journal of Medicine}, vol. 375,
  no.~13, pp. 1303--1305, 2016.

\bibitem{fehr2015coronaviruses}
A.~R. Fehr and S.~Perlman, ``Coronaviruses: an overview of their replication
  and pathogenesis,'' in \emph{Coronaviruses}.\hskip 1em plus 0.5em minus
  0.4em\relax Springer, 2015, pp. 1--23.

\bibitem{al2013middle}
S.~Al~Hajjar, Z.~A. Memish, and K.~McIntosh, ``Middle east respiratory syndrome
  coronavirus ({MERS-CoV}): a perpetual challenge,'' \emph{Annals of Saudi
  medicine}, vol.~33, no.~5, pp. 427--436, 2013.

\bibitem{Towner2004}
J.~S. Towner, P.~E. Rollin, D.~G. Bausch, A.~Sanchez, S.~M. Crary, M.~Vincent,
  W.~F. Lee, C.~F. Spiropoulou, T.~G. Ksiazek, M.~Lukwiya \emph{et~al.},
  ``Rapid diagnosis of {Ebola} hemorrhagic fever by reverse transcription-{PCR}
  in an outbreak setting and assessment of patient viral load as a predictor of
  outcome,'' \emph{Journal of Virology}, vol.~78, no.~8, pp. 4330--4341, 2004.

\bibitem{deLaVega2015}
M.-A. de~La~Vega, G.~Caleo, J.~Audet, X.~Qiu, R.~A. Kozak, J.~I. Brooks,
  S.~Kern, A.~Wolz, A.~Sprecher, J.~Greig \emph{et~al.}, ``{Ebola} viral load
  at diagnosis associates with patient outcome and outbreak evolution,''
  \emph{The Journal of Clinical Investigation}, vol. 125, no.~12, pp.
  4421--4428, 12 2015.

\bibitem{Ben-Shachar2016}
R.~Ben-Shachar, S.~Schmidler, and K.~Koelle, ``Drivers of inter-individual
  variation in {Dengue} viral load dynamics,'' \emph{PLOS Computational
  Biology}, vol.~12, no.~11, pp. 1--26, 11 2016.

\bibitem{Wang2003}
W.-K. Wang, D.-Y. Chao, C.-L. Kao, H.-C. Wu, Y.-C. Liu, C.-M. Li, S.-C. Lin,
  S.-T. Ho, J.-H. Huang, and C.-C. King, ``High levels of plasma {Dengue} viral
  load during defervescence in patients with {Dengue Hemorrhagic Fever}:
  Implications for pathogenesis,'' \emph{Virology}, vol. 305, no.~2, pp. 330 --
  338, 2003.

\bibitem{Wang2006}
W.~K. Wang, H.~L. Chen, C.~F. Yang, S.~C. Hsieh, C.~C. Juan, S.~M. Chang, C.~C.
  Yu, L.~H. Lin, J.~H. Huang, and C.~C. King, ``Slower rates of clearance of
  viral load and virus-containing immune complexes in patients with {Dengue
  hemorrhagic fever},'' \emph{Clin Infect Dis}, vol.~43, no.~8, pp. 1023--30,
  2006.

\bibitem{Mansuy2016}
J.~M. Mansuy, M.~Dutertre, C.~Mengelle, C.~Fourcade, B.~Marchou, P.~Delobel,
  J.~Izopet, and G.~Martin-Blondel, ``Zika virus: high infectious viral load in
  semen, a new sexually transmitted pathogen?'' \emph{The Lancet Infectious
  Diseases}, vol.~16, no.~4, p. 405, 2016.

\bibitem{deLaval2017}
F.~de~Laval, S.~Matheus, T.~Labrousse, A.~Enfissi, D.~Rousset, and S.~Briolant,
  ``Kinetics of {Zika} viral load in semen,'' \emph{New England Journal of
  Medicine}, vol. 377, no.~7, pp. 697--699, 2017.

\bibitem{Fourcade2016}
C.~Fourcade, J.-M. Mansuy, M.~Dutertre, M.~Delpech, B.~Marchou, P.~Delobel,
  J.~Izopet, and G.~Martin-Blondel, ``Viral load kinetics of {Zika} virus in
  plasma, urine and saliva in a couple returning from {Martinique, French West
  Indies},'' \emph{Journal of Clinical Virology}, vol.~82, pp. 1 -- 4, 2016.

\bibitem{Lerat1998}
H.~Lerat, S.~Rumin, F.~Habersetzer, F.~Berby, M.~A. Trabaud, C.~Trépo, and
  G.~Inchauspé, ``In vivo tropism of hepatitis {C} virus genomic sequences in
  hematopoietic cells: influence of viral load, viral genotype, and cell
  phenotype,'' \emph{Blood}, vol.~91, no.~10, pp. 3841--3849, 1998.

\bibitem{VABRET2020}
N.~Vabret, G.~J. Britton, C.~Gruber, S.~Hegde, J.~Kim, M.~Kuksin,
  R.~Levantovsky, L.~Malle, A.~Moreira, M.~D. Park \emph{et~al.}, ``Immunology
  of {COVID-19}: Current state of the science,'' \emph{Immunity}, vol.~52,
  no.~6, pp. 910 -- 941, 2020.

\bibitem{Tay2020}
M.~Z. Tay, C.~M. Poh, L.~Rénia, P.~A. MacAry, and L.~F.~P. Ng, ``The trinity
  of {COVID-19}: immunity, inflammation and intervention,'' \emph{Nature
  Reviews Immunology}, vol.~20, no.~6, pp. 363--374, 2020.

\bibitem{Faure2014}
E.~Faure, J.~Poissy, A.~Goffard, C.~Fournier, E.~Kipnis, M.~Titecat,
  P.~Bortolotti, L.~Martinez, S.~Dubucquoi, R.~Dessein \emph{et~al.},
  ``Distinct immune response in two {MERS-CoV}-infected patients: can we go
  from bench to bedside?'' \emph{PLoS One}, vol.~9, no.~2, p. e88716, 2014.

\bibitem{Baize2002}
S.~Baize, E.~M. Leroy, A.~J. Georges, M.~C. Georges-Courbot, M.~Capron,
  I.~Bedjabaga, J.~Lansoud-Soukate, and E.~Mavoungou, ``Inflammatory responses
  in {Ebola} virus-infected patients,'' \emph{Clinical and experimental
  immunology}, vol. 128, no.~1, pp. 163--168, 2002.

\bibitem{Levine2019}
M.~M. Levine, ``Monoclonal antibody therapy for {Ebola} virus disease,''
  \emph{New England Journal of Medicine}, vol. 381, no.~24, pp. 2365--2366,
  2019.

\bibitem{Fares-Gusmao2019}
R.~Fares-Gusmao, B.~C. Rocha, E.~Sippert, M.~C. Lanteri, G.~Áñez, and
  M.~Rios, ``Differential pattern of soluble immune markers in asymptomatic
  {Dengue}, {West Nile} and {Zika} virus infections,'' \emph{Scientific
  Reports}, vol.~9, no.~1, p. 17172, 2019.

\bibitem{Guzman2010}
M.~G. Guzman, S.~B. Halstead, H.~Artsob, P.~Buchy, J.~Farrar, D.~J. Gubler,
  E.~Hunsperger, A.~Kroeger, H.~S. Margolis, E.~Martínez \emph{et~al.},
  ``Dengue: a continuing global threat,'' \emph{Nature Reviews Microbiology},
  vol.~8, no.~12, pp. S7--S16, 2010.

\bibitem{Collins2019}
M.~H. Collins, H.~A. Tu, C.~Gimblet-Ochieng, G.~A. Liou, R.~S. Jadi, S.~W.
  Metz, A.~Thomas, B.~D. McElvany, E.~Davidson, B.~J. Doranz \emph{et~al.},
  ``Human antibody response to {Zika} targets type-specific quaternary
  structure epitopes,'' \emph{JCI Insight}, vol.~4, no.~8, 2019.

\bibitem{Klenerman2012}
P.~Klenerman and R.~Thimme, ``T cell responses in hepatitis {C}: the good, the
  bad and the unconventional,'' \emph{Gut}, vol.~61, no.~8, pp. 1226--34, 2012.

\bibitem{Hetta2014}
H.~Hetta, M.~Mehta, and M.~Shata, ``Gut immune response in the presence of
  hepatitis {C} virus infection,'' \emph{World Journal of Immunology}, vol.~4,
  no.~2, pp. 52--62, 2014.

\bibitem{Swann2016}
R.~E. Swann, V.~M. Cowton, M.~W. Robinson, S.~J. Cole, S.~T. Barclay, P.~R.
  Mills, E.~C. Thomson, J.~McLauchlan, and A.~H. Patel, ``Broad anti-hepatitis
  {C} virus {(HCV)} antibody responses are associated with improved clinical
  disease parameters in chronic {HCV} infection,'' \emph{Journal of Virology},
  vol.~90, no.~9, pp. 4530--4543, 2016.

\bibitem{BALASHAZY199115}
I.~Balásházy, W.~Hofmann, and T.~B. Martonen, ``Inspiratory particle
  deposition in airway bifurcation models,'' \emph{Journal of Aerosol Science},
  vol.~22, no.~1, pp. 15 -- 30, 1991.

\bibitem{Olufsen2000}
M.~S. Olufsen, C.~S. Peskin, W.~Y. Kim, E.~M. Pedersen, A.~Nadim, and
  J.~Larsen, ``Numerical simulation and experimental validation of blood flow
  in arteries with structured-tree outflow conditions,'' \emph{Annals of
  Biomedical Engineering}, vol.~28, no.~11, pp. 1281--1299, 2000.

\bibitem{Xu2020}
H.~Xu, L.~Zhong, J.~Deng, J.~Peng, H.~Dan, X.~Zeng, T.~Li, and Q.~Chen, ``High
  expression of {ACE2} receptor of {2019-nCoV} on the epithelial cells of oral
  mucosa,'' \emph{International Journal of Oral Science}, vol.~12, no.~1, p.~8,
  2020.

\bibitem{Vargas2020}
E.~A. Hernandez~Vargas and J.~X. Velasco-Hernandez, ``In-host modelling of
  {COVID-19} kinetics in humans,'' \emph{medRxiv}, 2020.

\bibitem{Wolfel2020}
R.~Wölfel, V.~M. Corman, W.~Guggemos, M.~Seilmaier, S.~Zange, M.~A. Müller,
  D.~Niemeyer, T.~C. Jones, P.~Vollmar, C.~Rothe \emph{et~al.}, ``Virological
  assessment of hospitalized patients with {COVID}-2019,'' \emph{Nature}, vol.
  581, no. 7809, pp. 465--469, 2020.

\bibitem{Zhang2004}
Y.~Zhang, J.~Li, Y.~Zhan, L.~Wu, X.~Yu, W.~Zhang, L.~Ye, S.~Xu, R.~Sun,
  Y.~Wang, and J.~Lou, ``Analysis of serum cytokines in patients with severe
  acute respiratory syndrome,'' \emph{Infection and immunity}, vol.~72, no.~8,
  pp. 4410--4415, 2004.

\bibitem{Cheung2005}
C.~Y. Cheung, L.~L.~M. Poon, I.~H.~Y. Ng, W.~Luk, S.-F. Sia, M.~H.~S. Wu, K.-H.
  Chan, K.-Y. Yuen, S.~Gordon, Y.~Guan, and J.~S.~M. Peiris, ``Cytokine
  responses in severe acute respiratory syndrome coronavirus-infected
  macrophages in vitro: possible relevance to pathogenesis,'' \emph{Journal of
  virology}, vol.~79, no.~12, pp. 7819--7826, 2005.

\bibitem{Jiang2005}
Y.~Jiang, J.~Xu, C.~Zhou, Z.~Wu, S.~Zhong, J.~Liu, W.~Luo, T.~Chen, Q.~Qin, and
  P.~Deng, ``Characterization of cytokine/chemokine profiles of severe acute
  respiratory syndrome,'' \emph{Am J Respir Crit Care Med}, vol. 171, no.~8,
  pp. 850--7, 2005.

\bibitem{barros2020engineering}
M.~T. Barros, P.~Doan, M.~Kandhavelu, B.~Jennings, and S.~Balasubramaniam,
  ``Engineering calcium signaling of astrocytes for neural-molecular computing
  logic gates,'' \emph{arXiv preprint arXiv:2007.06646}, 2020.

\bibitem{nakano2020roles}
T.~Nakano, Y.~Okaie, Y.~Kinugasa, T.~Koujin, T.~Suda, Y.~Hiraoka, and
  T.~Haraguchi, ``Roles of remote and contact forces in epithelial cell
  structure formation,'' \emph{Biophysical Journal}, 2020.

\bibitem{grebenstein2019molecular}
L.~Grebenstein, J.~Kirchner, W.~Wicke, A.~Ahmadzadeh, V.~Jamali, G.~Fischer,
  R.~Weigel, A.~Burkovski, and R.~Schober, ``A molecular communication testbed
  based on proton pumping bacteria: Methods and data,'' \emph{IEEE Transactions
  on Molecular, Biological and Multi-Scale Communications}, vol.~5, no.~1, pp.
  56--62, 2019.

\bibitem{liu2017using}
Y.~Liu, C.-Y. Tsao, E.~Kim, T.~Tschirhart, J.~L. Terrell, W.~E. Bentley, and
  G.~F. Payne, ``Using a redox modality to connect synthetic biology to
  electronics: Hydrogel-based chemo-electro signal transduction for molecular
  communication,'' \emph{Advanced healthcare materials}, vol.~6, no.~1, p.
  1600908, 2017.

\bibitem{balasubramaniam2011development}
S.~Balasubramaniam, N.~T. Boyle, A.~Della-Chiesa, F.~Walsh, A.~Mardinoglu,
  D.~Botvich, and A.~Prina-Mello, ``Development of artificial neuronal networks
  for molecular communication,'' \emph{Nano Communication Networks}, vol.~2,
  no. 2-3, pp. 150--160, 2011.

\bibitem{mcguiness2018experimental}
D.~T. McGuiness, S.~Giannoukos, A.~Marshall, and S.~Taylor, ``Experimental
  results on the open-air transmission of macro-molecular communication using
  membrane inlet mass spectrometry,'' \emph{IEEE Communications Letters},
  vol.~22, no.~12, pp. 2567--2570, 2018.

\bibitem{felicetti2014modeling}
L.~Felicetti, M.~Femminella, G.~Reali, P.~Gresele, M.~Malvestiti, and J.~N.
  Daigle, ``Modeling {CD40-based} molecular communications in blood vessels,''
  \emph{IEEE transactions on nanobioscience}, vol.~13, no.~3, pp. 230--243,
  2014.

\bibitem{vishveshwara2009intra}
S.~Vishveshwara, A.~Ghosh, and P.~Hansia, ``Intra and inter-molecular
  communications through protein structure network,'' \emph{Current Protein and
  Peptide Science}, vol.~10, no.~2, pp. 146--160, 2009.

\bibitem{lee2004isolation}
O.~K. Lee, T.~K. Kuo, W.-M. Chen, K.-D. Lee, S.-L. Hsieh, and T.-H. Chen,
  ``Isolation of multipotent mesenchymal stem cells from umbilical cord
  blood,'' \emph{Blood}, vol. 103, no.~5, pp. 1669--1675, 2004.

\bibitem{sottile2002stem}
V.~Sottile, C.~Halleux, F.~Bassilana, H.~Keller, and K.~Seuwen, ``Stem cell
  characteristics of human trabecular bone-derived cells,'' \emph{Bone},
  vol.~30, no.~5, pp. 699--704, 2002.

\bibitem{de2001multipotent}
C.~De~Bari, F.~Dell'Accio, P.~Tylzanowski, and F.~P. Luyten, ``Multipotent
  mesenchymal stem cells from adult human synovial membrane,'' \emph{Arthritis
  \& Rheumatism}, vol.~44, no.~8, pp. 1928--1942, 2001.

\bibitem{sato2005human}
Y.~Sato, H.~Araki, J.~Kato, K.~Nakamura, Y.~Kawano, M.~Kobune, T.~Sato,
  K.~Miyanishi, T.~Takayama, M.~Takahashi \emph{et~al.}, ``Human mesenchymal
  stem cells xenografted directly to rat liver are differentiated into human
  hepatocytes without fusion,'' \emph{Blood}, vol. 106, no.~2, pp. 756--763,
  2005.

\bibitem{gupta2020mesenchymal}
S.~Gupta, V.~Krishnakumar, Y.~Sharma, A.~K. Dinda, and S.~Mohanty,
  ``Mesenchymal stem cell derived exosomes: a nano platform for therapeutics
  and drug delivery in combating {COVID-19},'' \emph{Stem cell reviews and
  reports}, pp. 1--11, 2020.

\bibitem{islam2020perspective}
M.~T. Islam, M.~Nasiruddin, I.~N. Khan, S.~K. Mishra, M.~Kudrat-E-Zahan, T.~A.
  Riaz, E.~S. Ali, M.~S. Rahman, M.~S. Mubarak, M.~Martorell \emph{et~al.}, ``A
  perspective on emerging therapeutic interventions for {COVID-19},''
  \emph{Frontiers in public health}, vol.~8, 2020.

\bibitem{cruz2019preclinical}
T.~Cruz and M.~Rojas, ``Preclinical evidence for the role of stem/stromal cells
  in targeting ards,'' in \emph{Stem Cell-Based Therapy for Lung
  Disease}.\hskip 1em plus 0.5em minus 0.4em\relax Springer, 2019, pp.
  199--217.

\bibitem{zanoni2019role}
M.~Zanoni, M.~Cortesi, A.~Zamagni, and A.~Tesei, ``The role of mesenchymal stem
  cells in radiation-induced lung fibrosis,'' \emph{International journal of
  molecular sciences}, vol.~20, no.~16, p. 3876, 2019.

\bibitem{perestrelo2015microfluidic}
A.~R. Perestrelo, A.~C. {\'A}guas, A.~Rainer, and G.~Forte, ``Microfluidic
  organ/body-on-a-chip devices at the convergence of biology and
  microengineering,'' \emph{Sensors}, vol.~15, no.~12, pp. 31\,142--31\,170,
  2015.

\bibitem{vsiljak2020evolving}
H.~{\v{S}}iljak, N.~Ashraf, M.~T. Barros, D.~P. Martins, B.~Butler, A.~Farhang,
  N.~Marchetti, and S.~Balasubramaniam, ``Evolving intelligent reflector
  surface towards {6G} for public health: Application in airborne virus
  detection,'' \emph{arXiv preprint arXiv:2009.02224}, 2020.

\bibitem{tang2020human}
H.~Tang, Y.~Abouleila, L.~Si, A.~M. Ortega-Prieto, C.~L. Mummery, D.~E. Ingber,
  and A.~Mashaghi, ``Human organs-on-chips for virology,'' \emph{Trends in
  Microbiology}, 2020.

\bibitem{ringehan2017viral}
M.~Ringehan, J.~A. McKeating, and U.~Protzer, ``Viral hepatitis and liver
  cancer,'' \emph{Philosophical Transactions of the Royal Society B: Biological
  Sciences}, vol. 372, no. 1732, p. 20160274, 2017.

\bibitem{villenave2017human}
R.~Villenave, S.~Q. Wales, T.~Hamkins-Indik, E.~Papafragkou, J.~C. Weaver,
  T.~C. Ferrante, A.~Bahinski, C.~A. Elkins, M.~Kulka, and D.~E. Ingber,
  ``Human gut-on-a-chip supports polarized infection of coxsackie {B1} virus in
  vitro,'' \emph{PloS one}, vol.~12, no.~2, p. e0169412, 2017.

\bibitem{cifuente2011molecular}
J.~O. Cifuente, M.~F. Ferrer, C.~J. de~Giusti, W.-C. Song, V.~Romanowski, S.~L.
  Hafenstein, and R.~M. Gomez, ``Molecular determinants of disease in
  coxsackievirus {B1} murine infection,'' \emph{Journal of medical virology},
  vol.~83, no.~9, pp. 1571--1581, 2011.

\bibitem{johnson20163d}
B.~N. Johnson, K.~Z. Lancaster, I.~B. Hogue, F.~Meng, Y.~L. Kong, L.~W.
  Enquist, and M.~C. McAlpine, ``{3D} printed nervous system on a chip,''
  \emph{Lab on a Chip}, vol.~16, no.~8, pp. 1393--1400, 2016.

\bibitem{benam2016small}
K.~H. Benam, R.~Villenave, C.~Lucchesi, A.~Varone, C.~Hubeau, H.-H. Lee, S.~E.
  Alves, M.~Salmon, T.~C. Ferrante, J.~C. Weaver \emph{et~al.}, ``Small
  airway-on-a-chip enables analysis of human lung inflammation and drug
  responses in vitro,'' \emph{Nature methods}, vol.~13, no.~2, pp. 151--157,
  2016.

\bibitem{si2019discovery}
L.~Si, R.~Prantil-Baun, K.~H. Benam, H.~Bai, M.~Rodas, M.~Burt, and D.~E.
  Ingber, ``Discovery of influenza drug resistance mutations and host
  therapeutic targets using a human airway chip,'' \emph{bioRxiv}, p. 685552,
  2019.

\bibitem{deng2017microfluidic}
Y.~Deng, M.~Pierobon, and A.~Nallanathan, ``A microfluidic feed forward loop
  pulse generator for molecular communication,'' in \emph{GLOBECOM 2017-2017
  IEEE Global Communications Conference}.\hskip 1em plus 0.5em minus
  0.4em\relax IEEE, 2017, pp. 1--7.

\bibitem{hamidovic2019passive}
M.~Hamidovi{\'c}, W.~Haselmayr, A.~Grimmer, R.~Wille, and A.~Springer,
  ``Passive droplet control in microfluidic networks: A survey and new
  perspectives on their practical realization,'' \emph{Nano Communication
  Networks}, vol.~19, pp. 33--46, 2019.

\bibitem{fronczek2015biosensors}
C.~F. Fronczek and J.-Y. Yoon, ``Biosensors for monitoring airborne
  pathogens,'' \emph{Journal of laboratory automation}, vol.~20, no.~4, pp.
  390--410, 2015.

\bibitem{grebenstein2018biological}
L.~Grebenstein, J.~Kirchner, R.~S. Peixoto, W.~Zimmermann, F.~Irnstorfer,
  W.~Wicke, A.~Ahmadzadeh, V.~Jamali, G.~Fischer, R.~Weigel \emph{et~al.},
  ``Biological optical-to-chemical signal conversion interface: A small-scale
  modulator for molecular communications,'' \emph{IEEE transactions on
  nanobioscience}, vol.~18, no.~1, pp. 31--42, 2018.

\bibitem{mcguiness2018parameter}
D.~T. McGuiness, S.~Giannoukos, A.~Marshall, and S.~Taylor, ``Parameter
  analysis in macro-scale molecular communications using advection-diffusion,''
  \emph{IEEE Access}, vol.~6, pp. 46\,706--46\,717, 2018.

\bibitem{roumpeka2017review}
D.~D. Roumpeka, R.~J. Wallace, F.~Escalettes, I.~Fotheringham, and M.~Watson,
  ``A review of bioinformatics tools for bio-prospecting from metagenomic
  sequence data,'' \emph{Frontiers in genetics}, vol.~8, p.~23, 2017.

\bibitem{soria2015overview}
R.~E. Soria-Guerra, R.~Nieto-Gomez, D.~O. Govea-Alonso, and S.~Rosales-Mendoza,
  ``An overview of bioinformatics tools for epitope prediction: implications on
  vaccine development,'' \emph{Journal of biomedical informatics}, vol.~53, pp.
  405--414, 2015.

\bibitem{garrido2008evaluation}
C.~Garrido, V.~Roulet, N.~Chueca, E.~Poveda, A.~Aguilera, K.~Skrabal,
  N.~Zahonero, S.~Carlos, F.~Garc{\'\i}a, J.~L. Faudon \emph{et~al.},
  ``Evaluation of eight different bioinformatics tools to predict viral tropism
  in different human immunodeficiency virus type 1 subtypes,'' \emph{Journal of
  clinical microbiology}, vol.~46, no.~3, pp. 887--891, 2008.

\bibitem{wang2015xander}
Q.~Wang, J.~A. Fish, M.~Gilman, Y.~Sun, C.~T. Brown, J.~M. Tiedje, and J.~R.
  Cole, ``Xander: employing a novel method for efficient gene-targeted
  metagenomic assembly,'' \emph{Microbiome}, vol.~3, no.~1, p.~32, 2015.

\bibitem{balasubramaniam2013multi}
S.~Balasubramaniam and P.~Li{\`o}, ``Multi-hop conjugation based bacteria
  nanonetworks,'' \emph{IEEE Transactions on nanobioscience}, vol.~12, no.~1,
  pp. 47--59, 2013.

\bibitem{bilgin2018dna}
B.~A. Bilgin, E.~Dinc, and O.~B. Akan, ``{DNA-based} molecular
  communications,'' \emph{IEEE Access}, vol.~6, pp. 73\,119--73\,129, 2018.

\bibitem{veletic2019information}
M.~Veleti{\'c} and I.~Balasingham, ``An information theory of
  neuro-transmission in multiple-access synaptic channels,'' \emph{IEEE
  Transactions on Communications}, vol.~68, no.~2, pp. 841--853, 2019.

\bibitem{barros2015comparative}
M.~T. Barros, S.~Balasubramaniam, and B.~Jennings, ``Comparative end-to-end
  analysis of {Ca}2+-signaling-based molecular communication in biological
  tissues,'' \emph{IEEE Transactions on Communications}, vol.~63, no.~12, pp.
  5128--5142, 2015.

\bibitem{barros2017ca2+}
M.~T. Barros, ``Ca2+-signaling-based molecular communication systems: Design
  and future research directions,'' \emph{Nano Communication Networks},
  vol.~11, pp. 103--113, 2017.

\bibitem{sompayrac2019immune}
L.~M. Sompayrac, \emph{How the immune system works}.\hskip 1em plus 0.5em minus
  0.4em\relax John Wiley \& Sons, 2019.

\end{thebibliography}



\end{document}